\documentclass[preprints,article,accept,pdftex,oneauthor]{Definitions/mdpi} 

\firstpage{1} 
\pubvolume{1}
\issuenum{1}
\articlenumber{0}
\pubyear{2023}
\copyrightyear{2023}
\externaleditor{Academic Editor: Firstname Lastname}
\datereceived{3 October 2023} 
\daterevised{23 October 2023} 
\dateaccepted{25 October 2023 } 
\datepublished{ } 
\hreflink{https://doi.org/} 
\pdfoutput=1 

\usepackage{braket}
\newcommand{\prd}{Phys. Rev. D}
\newcommand{\prl}{Phys. Rev. Lett.}
\newcommand{\jcap}{J. Cosmol. Astropart. Phys.}

\Title{Exploring Neutrino Mass Orderings through Supernova Neutrino Detection}

\TitleCitation{Exploring Neutrino Mass Orderings through Supernova Neutrino Detection}

\Author{Maria Manuela Saez $^{1,2}$\orcidA{}}

\AuthorNames{\highlighting{Maria} Manuela Saez}

\AuthorCitation{Saez, M.M.}

\address{%
$^{1}$ \quad Department of Physics, University of California, Berkeley, CA 94720, USA; manuela.saez@berkeley.edu\\
$^{2}$ \quad \textls[-10] RIKEN Interdisciplinary Theoretical and Mathematical Sciences Program (iTHEMS), 2-1 Hirosawa, Wako, Saitama 351-0198, Japan
 }

\abstract{Core-collapse supernovae (SNe) are one of the most powerful cosmic sources of neutrinos, with energies of several MeV. The emission of neutrinos and antineutrinos of all flavors carries away the gravitational binding energy of the compact remnant and drives its evolution from the hot initial to the cold final states. Detecting these neutrinos from Earth and analyzing the emitted signals present a unique opportunity to explore the neutrino mass ordering problem. This research outlines the detection of neutrinos from SNe and their relevance in understanding the neutrino mass ordering. The focus is on developing a model-independent analysis strategy, achieved by comparing distinct detection channels in large underground detectors. The objective is to identify potential indicators of mass ordering within the neutrino sector.
Additionally, a thorough statistical analysis is performed on the anticipated neutrino signals for both mass orderings. Despite uncertainties in supernova explosion parameters, an exploration of the parameter space reveals an extensive array of models with significant sensitivity to differentiate between mass orderings. The assessment of various observables and their combinations underscores the potential of forthcoming supernova observations in addressing the neutrino mass ordering problem.}


\begin{document}


\section{Introduction}
Core-collapse supernovae (SNe) are the final evolutionary stage of stars with masses $M \gtrsim 8M_\odot$ and represent a long-awaited observation target for neutrino telescopes. To explain these events, interdisciplinary research that combines nuclear physics, particle physics, and astrophysics is needed.
Neutrinos are important for studying the energy balance involved in SN collapses, because only about $1\%$ of the gravitational binding energy is released as kinetic energy in the compact object formation, while the remaining $99\%$ is carried out by neutrinos of all flavors, with energies of several MeV \cite{Woosley:2002}. The mechanisms leading to neutrino production in the SN core are, mainly, electron capture by nucleons $e^{-}+p\rightarrow n+\nu_e$, pair annihilation  $e^+ + e^- \rightarrow \nu_e+\bar{\nu_e}$, flavor conversion $\nu_e+\bar{\nu_e}\rightarrow \nu_{\tau,\mu}+\bar{\nu}_{\tau,\mu}$, and  nucleon bremsstrahlung $N+N\rightarrow N'+N'+\nu+\bar{\nu}$ \cite{Janka:2016}. Once they have traveled through the stellar material and space, neutrinos reaching the Earth can be detected, providing precious information on the stellar core \cite{Mirizzi:2016}.

Studying the signals that the neutrinos leave in the detectors, with an effective neutrino flavor discrimination, it is possible to infer properties on their physics, because the structure of the neutrino mass spectrum and lepton mixing is imprinted into the detected signal.
Neutrinos were already observed for the 1987A SN in the Large Magellanic Cloud. During this event, two water Cherenkov detectors, Kamiokande-II and the Irvine--Michigan--Brookhaven (IMB) experiment, observed neutrino interaction events at a time consistent with the estimated time of the collapse \cite{Hirata:1987,Bionta:1987}. Two scintillator detectors, Baksan and LSD~\cite{Alekseev:1987,Aglietta:1987}, also reported observations; the latter report was controversial because the events were recorded several hours early. At present, several detectors are ready and waiting for the detection of SN neutrinos from the next galactic explosion. SN neutrinos can be detected via electroweakly or strongly interacting products of weak charged-current (CC) and neutral-current (NC) interactions with electrons and nuclei. 
Some of the relevant interaction channels for current (and future) detectors are the inverse beta decay and neutrino--proton elastic scattering for scintillation detectors, the inverse beta decay and neutrino--electron elastic scattering for water Cherenkov detectors, and the absorption interaction on $^{40}$Ar in liquid Argon time projection chambers, among others.

Analytical models exist for the SN neutrino energy spectra that are useful for studying and predicting SN neutrino signals. The spectral distribution is often parametrized by a three-parameter fit that allows for deviations from a strictly thermal spectrum (motivated by analytic simplicity) \cite{2010:Hudepohl,Keil:2003}. These three parameters are the total neutrino energy $\epsilon_\alpha$, the neutrino mean energy $\braket{E_\alpha}$, and the spectral index or shape parameter $\beta_\alpha$, where $\alpha=(\nu_e, \bar{\nu}_e, \nu_x)$. 
SN simulations provide the indicative values of the mentioned parameters. However, they depend on many details of the neutrino transport inside the star, the properties of the incompletely known neutron-star equation of state, the properties of the collapsing star, and time \cite{OConnor:2018,Burrows:2021}. During the accretion phase, which takes place a few tens to
hundreds of milliseconds after the bounce, the expected neutrino energy spectrum
would exhibit a flavor hierarchy $\braket{E_{\nu_e}} < \braket{E_{\bar{\nu}_e}} < \braket{E_{\nu_x}}$ \cite{Janka:2016}. 
In recent years, several works have presented sensitivity studies of the energy spectrum of neutrinos concerning spectral parameters and developed strategies to minimize dependence on spectral models~\cite{vonkrosigk:2013,Li:2019}.
An interesting approach is through a combination of observables, for example, comparing CC- and NC-induced events \cite{Lunardini:2001v2,Lunardini:2003,Capozzi:2018}, to achieve model-independent signatures. Also, the ratio of events in these detection channels can distinguish between different neutrino mixing effects and processes. In Reference \cite{Capozzi:2018}, the authors study the possibility of disentangling a pure matter effect from a complete flavor equalization on the fluxes due to the \mbox{neutrino--neutrino interactions.}

The analysis and reconstruction of SN neutrino fluxes is an interesting tool to clarify the role of neutrinos in stellar explosion events and nucleosynthesis, as well as for studying physics at high densities and, in particular, the neutrino oscillation phenomena \cite{Mirizzi:2016,Tamborra:2012}. The neutrino oscillation phenomenon consists of a quantum process in which a flavor neutrino is described as a superposition of mass eigenstates, allowing it to change families between its emission and detection. Different detectors on Earth have measured neutrino fluxes from the sun, atmosphere, and reactors and reported the presence of this phenomenon consistently~\cite{Giunti:2003,Esteban:2020,Kajita:2010}.
The neutrino mass ordering is one of the major open issues in this regard, being one of the experimental priorities in the area of particle physics. This is called \emph{ {normal} 
} (\emph{ {inverted}}) if $\Delta m^{2}_{31} = m_3^2 -m_1^2 \geq 0\, (\leq 0)$.  Finding out which of the two cases occurs in nature has profound implications for cosmology, searches for neutrino mass, and studies of double-beta decays. In particular, neutrinos from SNe can be used to study this problem, because studying the neutronization burst, the time profile on the early signal, and the Earth matter effect can provide evidence of the neutrino mass ordering \cite{Scholberg:2018,Brdar:2022,Valls:2022}.

Given the aforementioned information, it is expected that a possible difficulty is that both core-collapse physics and neutrino physics affect the nature of the SN neutrino burst, and it may not be trivial to disentangle the two. The more experimental data we can gather in as many detectors around the globe as possible, and as far as we can find model-independent signatures, the better our chances will be of disentangling the various effects.

This research investigates whether the neutrino counts from SNe within the first second after the burst, detected across various channels and detectors, can elucidate the correct neutrino mass ordering. The main goal is to formulate a model-independent analysis approach by comparing diverse detection channels in large underground detectors, aiming to identify potential indicators of mass ordering in the neutrino sector. Furthermore, a thorough statistical analysis is conducted on the expected signals for both mass orderings, aiming to ascertain its discernibility.

\section{Supernova Neutrino Fluxes}\label{intro}

The time-integrated neutrino flux streaming off the SN can be expressed as (for the un-oscillated case): 

\begin{equation}
    {F^0_{\alpha} (E)}=\frac{\epsilon_{\alpha}}{4\pi d^2\braket{E_\alpha}}f_{\nu_{\alpha}}(E) \, , \label{flux}
\end{equation}
where $\braket{E_\alpha}$ are the neutrino
mean energies, $\epsilon_\alpha$ is the total neutrino energy (equivalent to the time-integrated luminosity) for which we assume equipartition among all the neutrino flavors, $d$ is the SN distance, and $f_{\nu_{\alpha}}(E)$ is the time-averaged energy spectrum  \cite{Keil:2003} given by   
\begin{equation}
    f_{\nu_{\alpha}}(E)=\frac{(1+\beta_\alpha)^{1+\beta_\alpha}}{\Gamma(1+\beta_\alpha)} \frac{E^{\beta_\alpha}}{\braket{E_\alpha}^{\beta_\alpha+1}}exp\left[{-(\beta_\alpha+1)}\frac{E}{\braket{E_\alpha}} \right] \label{distro-espectral}
\end{equation}
where $\beta$ is the pinching parameter of the spectral distribution. 
Given that the values of the spectral parameters depend on the SN simulation or model, the variation within specific ranges will be studied.

To calculate the neutrino fluxes including oscillation effects (${F_{\alpha} (E)}$), it is necessary to determine how the neutrino spectral distributions from Equation \ref{distro-espectral} are altered by the oscillations, by solving the flavor evolution equations.
Calling $\rho$ ($\bar{\rho}$) to the neutrino (antineutrino)-distribution function in its matrix form and $\mathcal{H}$ ($\bar{\mathcal{H}}$) the neutrino (antineutrino) Hamiltonian in the flavor basis, the differential equations that give the dependence of the neutrino (antineutrino)-distribution functions upon the radius are \cite{Balantekin:2004,Tamborra:2012}
\begin{eqnarray}\label{flavor-evol}
i\frac{\partial \rho}{\partial r}=\left[\mathcal{H},\rho\right]\nonumber \, ,\hspace{2cm}
i\frac{\partial \bar{\rho}}{\partial r}=\left[\bar{\mathcal{H}},\bar{\rho}\right]\, .
\end{eqnarray}

The Hamiltonian can be written as
\begin{eqnarray}
\mathcal{H}&=&\mathcal{H}^{vac}+\mathcal{H}^{m}+\mathcal{H}^{\nu-\nu} \, ,
\end{eqnarray}
where $\mathcal{H}^{vac}$ describes neutrino oscillations in vacuum, $\mathcal{H}^{m}$ represents the neutrino--matter interactions, and $\mathcal{H}^{\nu-\nu}$ takes into account the neutrino--neutrino interactions. 

\subsection*{Three-Active Scheme: The Two Flavor Approximations}\label{H-3active} 
 This work focuses on the first milliseconds until the first second of the SN event. In this stage, the effects of matter are dominant, and the self-induced effects can be neglected. Furthermore, during the whole accretion phase the matter potential is also expected to dominate over the neutrino--neutrino potential \cite{Sarikas:2011,Chakraborty:2011} (see appendix \ref{coll} for details and discussion).
 Operating within the three-active scheme, the calculation of SN neutrino fluxes incorporates both vacuum and matter effects \cite{Mirizzi:2016}. The mixing mechanism is assumed to remain unaffected by CP violations. Additionally, a rotation can be performed in this subspace to diagonalize the submatrix $\mu , \tau$ of \mbox{Equation (\ref{flavor-evol}) \cite{Dighe:2000,Tamborra:2012}.}
The $\nu_\mu$ and $\nu_\tau$ fluxes in an SN are very similar and these two flavors play symmetric roles.
Therefore, it is useful to define a linear combination $\nu_x$ that is essentially identical with the $m_3$
mass eigenstate and mixes with $\nu_e$ by means of the $\theta_{13}$ mixing angle \cite{Tamborra:2012,cottingham:2001}.
 Because the matter density (and therefore the potential) decreases with the star radius, active neutrinos exhibit two Mikheyev--Smirnov--Wolfenstein (MSW) resonances called H (high density) and L (low density) where the flavor conversion mechanism is amplified \cite{Dighe:2000}.  Instead of using the adiabatic approximation, we solved the complete equations numerically, considering the $\mathcal{H}^{vac}$ and $\mathcal{H}^{m}$ terms in the Hamiltonian (see \mbox{Appendix \ref{evol_eq} for details).}

For this work, the mixing parameters considered are from PDG 2020 \cite{Zyla:2020}:{\linebreak} $\sin^2(2\theta_{13})=0.0241$ and $\Delta m^2_{31}= 2.52 \times 10^{-3} eV^2$ ($\Delta m^2_{31}= -2.24 \times 10^{-3} eV^2$) for the normal (inverted) mass ordering.

\begin{figure} [H]
\includegraphics[width=1.\textwidth]{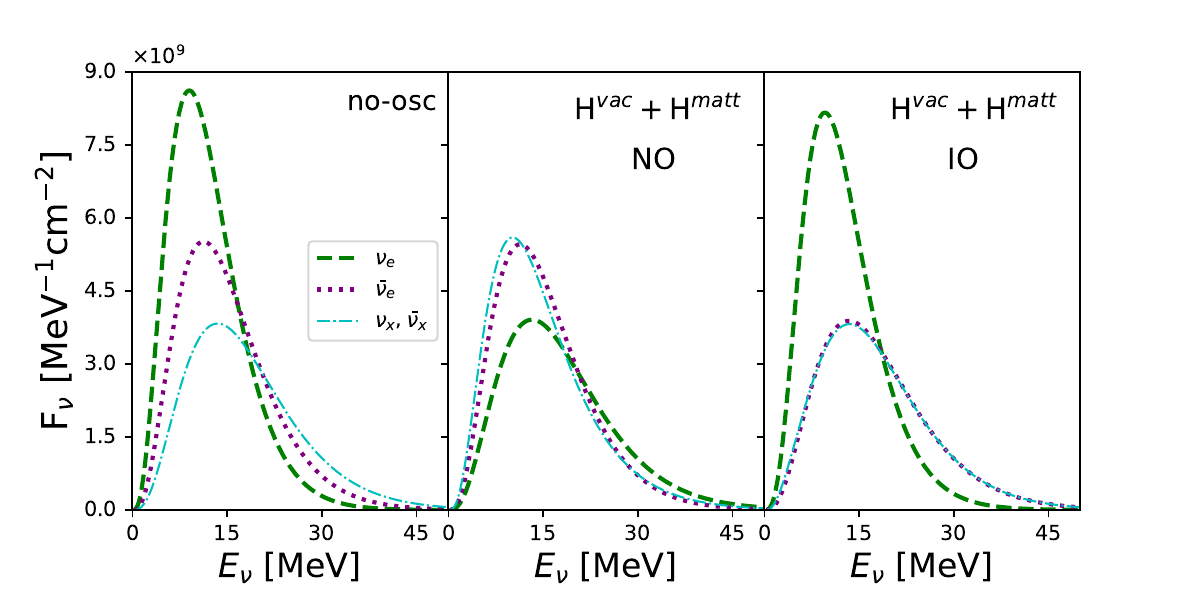}
\caption{ {
		SN neutrino} 
 fluence for the no-oscillation case (left panel), for the three-active mixing scheme including both vacuum and matter effects considering normal mass ordering (middle panel) and invert mass ordering (right panel). All the plots consider only the first second of the SN event, 
 which considers a total neutrino energy of 150foe. In this plot, we have set the spectral parameters to $\beta=3$, $\braket{E_{\nu_e}}=12$ MeV, $\braket{E_{\bar{\nu}_e}}=15$ MeV, and  $\braket{E_{\nu_x}}=18$  {MeV} 
 \cite{Dasgupta:2011}.}
\label{fluxes_SN_1st_sec}
\end{figure}

In Figure \ref{fluxes_SN_1st_sec}, the fluxes obtained following the above description for an SN event at 10~kpc of distance are presented. The non-oscillation case is shown in the first column, while in the second and third columns we show the fluxes for the case in which the neutrino oscillations for NO (2nd column) and IO (3rd column) are considered.

\section{Selected Detectors and Interactions Channels}\label{detectors}
To calculate the expected signal on Earth, the interactions of SN neutrinos in various channels have been computed. This involved selecting detectors of different technology and incorporating channels sensitive to different flavors.
In Table \ref{tab:selected_detectors_and_channels}, the chosen detectors, their host laboratory (location), the type of channel (neutral or charged current) they can detect, and the considered processes are presented.
\begin{table*}[ht]
\begin{center}
\caption{Selected detectors to perform the calculations of the expected signal. The selected interaction channels (type of current and processes) are shown in the last two columns.}
\label{tab:selected_detectors_and_channels}
{\renewcommand{\arraystretch}{1.5}
\begin{tabular}{ccccc}
\toprule
\textbf{Detector}&\textbf{ Location} & \textbf{Tot. Mass (Fid. Mass)} & \textbf{Current} & \textbf{Process}\\ \midrule
SNO$+$& SNOLAB & $\sim780$ t ($\sim 0.45$ kt) & CC & {\footnotesize $\bar{\nu_e}+p \rightarrow n +e^+$}  \\
& & & NC &{\footnotesize $\nu +p \rightarrow \nu'+p$} \\
& & & NC & {\footnotesize $\nu+^{12}\rm{C} \rightarrow ^{12}\rm{C}^{*}(15.11 \rm{MeV})$$+\nu'$}\\
\midrule
HALO& SNOLAB  & $\sim79$ t & CC & {\footnotesize $\nu_e +^{208}\rm{Pb} \rightarrow ^{207}\rm{Bi}+n +e^{-}$} \\
& & & CC & {\footnotesize $\nu_e +^{208}\rm{Pb} \rightarrow ^{206}\rm{Bi}+2n +e^{-}$}\\
& & & NC & {\footnotesize $\nu_x +^{208}\rm{Pb} \rightarrow ^{207}\rm{Pb}+n$}\\
& & & NC & {\footnotesize $\nu_x +^{206}\rm{Pb} \rightarrow ^{207}\rm{Pb}+2n$}\\ 
\midrule
DUNE& Fermilab & $\sim 70$ kt ($\sim40$ kt) & CC & {\footnotesize $\nu_e+^{40}\rm{Ar} \rightarrow e^- + ^{40}\rm{K}^*$} \\
& & & CC & {\footnotesize $\bar{\nu}_e+^{40}\rm{Ar} \rightarrow e^+ + ^{40}\rm{Cl}^*$} \\
& & & NC & {\footnotesize ${\nu}+^{40}\rm{Ar} \rightarrow$$ \nu + ^{40}\rm{Ar}^*$} \\
& & & NC + CC& {\footnotesize $\nu+e^- \rightarrow \nu+e^- $} \\
\bottomrule
\end{tabular}}
\end{center}
\end{table*}

\subsection{Interactions in SNO+}
\begin{figure}[h]
        \includegraphics[width=1.0\textwidth]{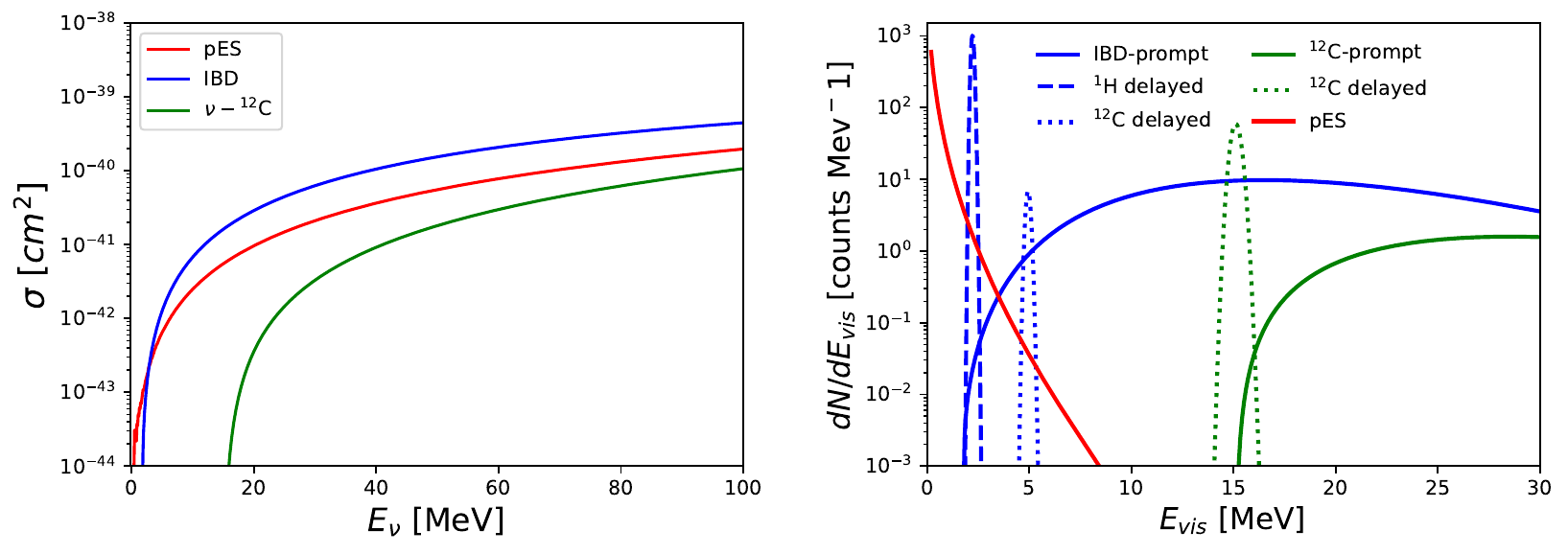}\caption{
        	Left panel: Cross sections as a function of the neutrino energy for the three SNO+ reactions listed in Table \ref{tab:selected_detectors_and_channels}. Right panel: Visible energy spectra for the IBD (blue), pES (red),  {and} $\nu-C^{12}$ (green) detection channels studied for the liquid scintillator SNO+. These events are induced by SN neutrinos following the fluence from Figure \ref{fluxes_SN_1st_sec}. } \label{total-SNO} 
    \end{figure}

The SNO+ experiment, located 2 km underground at SNOLAB in Sudbury, Canada,  is a liquid scintillator detector sensitive to neutrinos emitted from an SN in the Milky Way through the interaction channels mentioned in Table \ref{tab:selected_detectors_and_channels}.
Its detection efficiency is assumed to be perfect above the 200 keV threshold. Here, it is assumed that $N_p=3.32\times10^{31}$  free protons in a fiducial mass of 0.45 kt and an energy resolution of $5\%/\sqrt{E_{vis}}$ \cite{SNOplus:2015}. Neutrino interactions with free protons are considered through inverse beta decay (IBD) $\left(\bar{\nu}_e + p \rightarrow n + e^+\right)$, neutrino--proton elastic scattering (pES) $\nu +p \rightarrow \nu'+p$, and neutrino--nucleus reactions in $^{12}$C ($\nu-C^{12}$),  {given by} 
 $\nu + \rm{^{12}C} \rightarrow \rm{^{12}C}^* (15.11~\rm{MeV})$$\, +\,\nu'$. 
 The details of the calculation of the event rate and cross section are given in the Appendix \ref{channels}. The cross sections for the reactions considered for the SNO$+$ detector as a function of the neutrino energy are shown in the left panel of Figure \ref{total-SNO}, while the right panel shows the expected signal for all the mentioned channels. 

\subsection{Interactions in HALO}
The HALO detector is dedicated to the study of SN neutrinos and it is able to observe the neutrons emitted from electron neutrino scattering on lead from charged current (CC) and NC events. 
Measuring one- and two-neutron events on lead is particularly attractive to extract information about the SN neutrino temperatures \cite{Volpe:2011}.
For the listed reactions in \mbox{Table \ref{tab:selected_detectors_and_channels}}, neutrons are detected using ${^{3}He}$ counters as performed for the SNO experiment. 
HALO will measure both NC and CC events without distinguishing them because the outgoing electrons are not detected.
Because the outgoing lepton is not identified, the total event rate $N^{tot}$ is given by the sum of both the NC and CC expected rates
\begin{equation}
    N^{tot}_{{1n}(2n)}= N^{NC}_{{1n}(2n)}+ N^{CC}_{{1n}(2n)}
\end{equation}
where $N_{{1n}(2n)}$ refers to 1-neutron (1n) or 2-neutron (2n) event rates. 
The CC and NC event rates are given by 
\begin{eqnarray}
    N^{CC}_{{1n}(2n)}&=&N_{Pb}\int \frac{d\sigma^{CC}_{{1n}(2n)}}{dE} F_{\nu_e} dE \nonumber\\
    N^{NC}_{{1n}(2n)}&=&N_{Pb}\int \sum _{\alpha=e,\mu,\tau} \left[\frac{d\sigma^{NC, \nu}_{{1n}(2n)}}{dE} F_{\nu_\alpha}+\frac{d\sigma^{NC, \bar{\nu}}_{{1n}(2n)}}{dE} F_{\bar{\nu}_\alpha} \right]dE 
\end{eqnarray}

\begin{figure}[ht]
\includegraphics[width=1.0\textwidth]{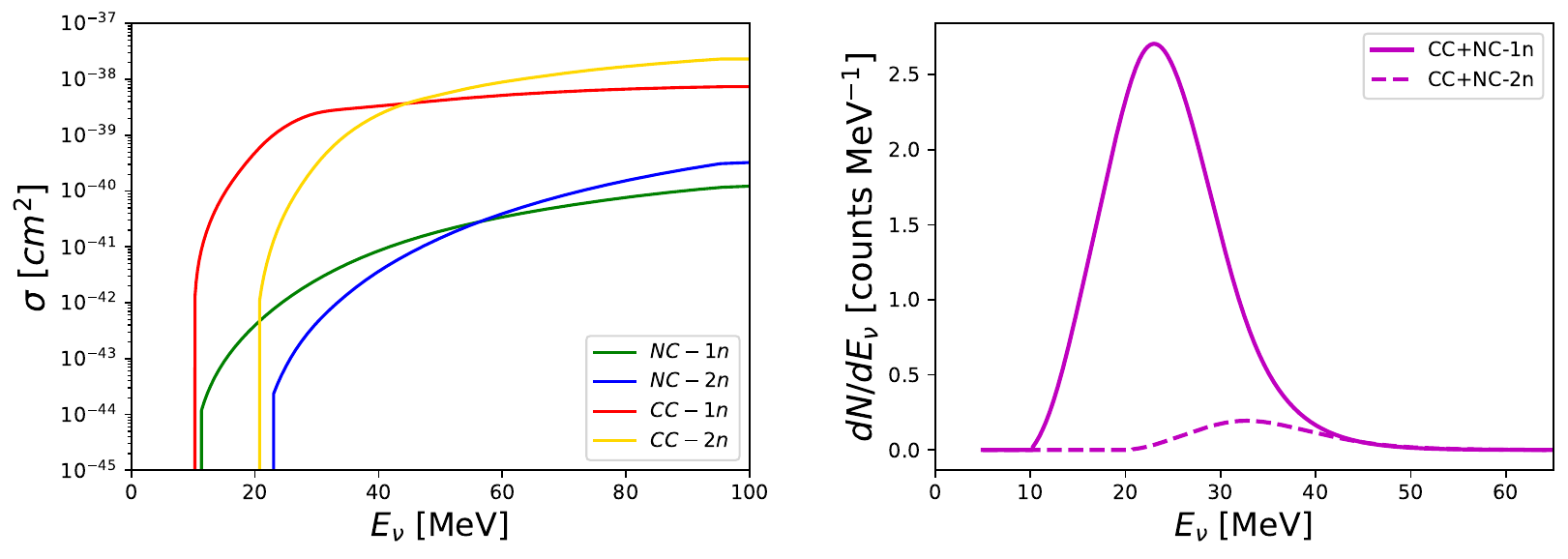}
\caption{
	Left panel: Cross sections for the four lead reactions listed in Table \ref{tab:selected_detectors_and_channels}. Right panel: Expected NC + CC $\nu-\rm{^{208}Pb}$ reactions for 1n and 2n emission for the HALO detector as a function of the neutrino energy for the full SN event. The SN neutrino fluence is assumed to follow results from Figure \ref{fluxes_SN_1st_sec} (here only the no-osc case is shown). }
\label{total-Pb208}
\end{figure}

The cross sections for the CC process were taken from Ref. \cite{Engel:2003}, 
while for the NC channels, the cross sections were extracted from SNOwGlobes \cite{Scholberg:2021}.  The cross sections as a function of the neutrino energy are plotted in the left panel of Figure \ref{total-Pb208}. The right panel shows the signal expected in the detector in both 1n and 2n cases by assuming a detector efficiency of 36$\%$ \cite{Volpe:2011}.  

\subsection{Interactions in DUNE}

\begin{figure}[ht]
\includegraphics[width=1.\textwidth]{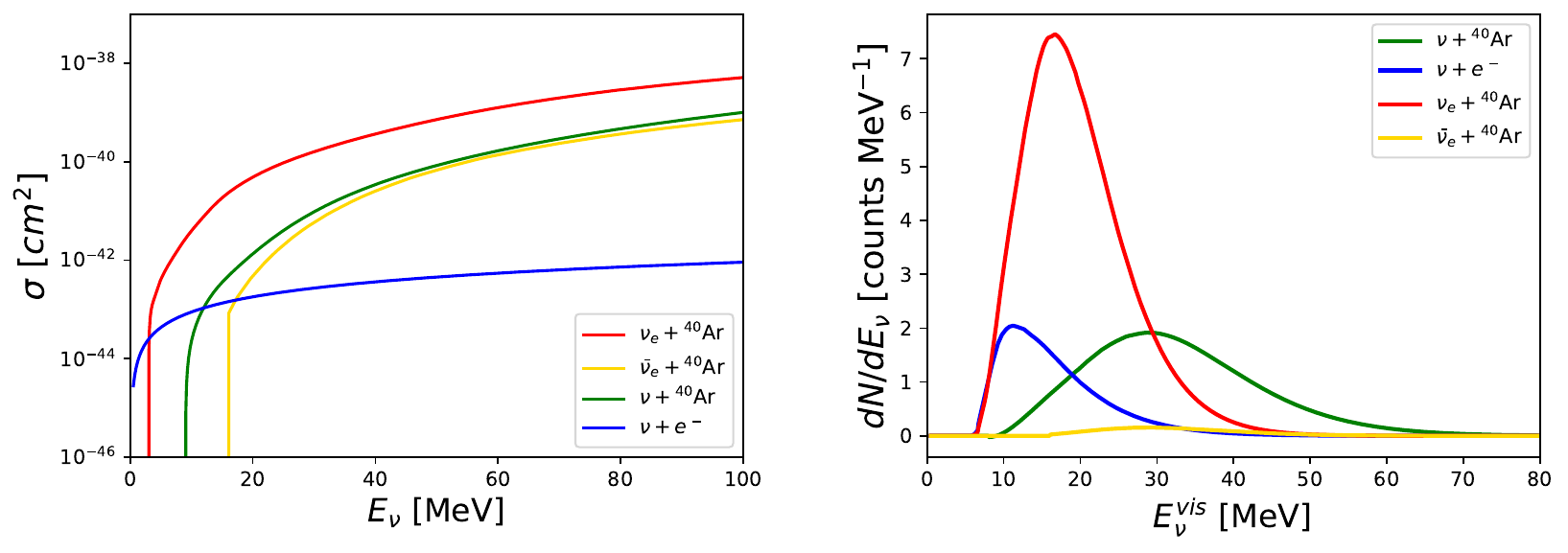}
\caption{
	Left panel: Cross sections for the four DUNE detector reactions listed in Table \ref{tab:selected_detectors_and_channels}. Right panel: Expected  NC + CC  signals for the DUNE detector as a function of the neutrino energy  for the 1st second of the SN event. The SN neutrino fluence is assumed to follow results from Figure \ref{fluxes_SN_1st_sec} (here only the no-osc case is shown). }
\label{total-Ar}
\end{figure}

The Deep Underground Neutrino Experiment (DUNE) will be made up of four 10 kton liquid argon time projection chambers (LArTPCs).  
DUNE’s dynamic range is such that it is also sensitive to neutrinos with energies down to about 5 MeV. CC interactions of neutrinos from around 5 MeV to several tens of MeV create short electron tracks in liquid argon, potentially accompanied by gamma ray and other secondary particle signatures. This regime is of particular interest for the  detection of neutrinos from a galactic SN \cite{Abi:2021}. 
DUNE will have a 70 kton liquid argon mass in total \cite{Abi:2020-TDR}, of which 40 kton will be fiducial mass (10 kton fiducial mass per module). For each module, a number of argon nuclei $\rm{N}_{Ar} = 1.5\times10^{32}$ and of electrons $\rm{N}_e=2.7\times10^{33}$ is assumed \cite{Abi:2020}. 
A detector efficiency, an energy bin distribution, and an energy resolution extracted from \cite{Abi:2021} have also been considered. In this case, the MARLEY~\cite{Gardiner:2021} cross sections implemented in SNOwGLoBES~\cite{Scholberg:2021} are used, as shown in the left panel of Figure \ref{total-Ar}. The right panel displays the expected signal \mbox{in the detector.}
\section{Combined Signals}\label{sec_ratios}  

The signatures of the neutrino mixing effects in the detected signals are closely related to the difference between the electronic and non-electronic fluxes. However, as mentioned before, the fluxes are model-dependent, posing a challenge in the analysis. To address this issue, two approaches have been employed: On the one hand, several works performed a global fit of the data, simultaneously determining the oscillation parameters and the parameters of the original fluxes \cite{Barger:2001,Minakata:2001,GalloRosso:2017}. On the other hand, other research has employed a combination of observables to reduce the reliance on specific models and the number of free parameters \cite{Capozzi:2018,Lunardini:2003,Engel:2002}. By utilizing this strategy, the analysis becomes less model dependent, providing a more robust characterization of the neutrino mixing phenomena.

Here, a hybrid approach is adopted. In addition to the regular counts expected for each interaction channel, combinations between them are studied. Specifically, the following ratios are analyzed:

\begin{enumerate}
\item The ratio of 1n and 2n events expected in the HALO detector ($\frac{\rm{N_{1n}}}{\rm{N_{2n}}}$).
\item The ratio between the pES and IBD events expected in the SNO+ detector ($\frac{\rm{N_{pES}}}{\rm{N_{IBD}}}$).
\item The ratio of CC and NC events on Argon in DUNE ($\frac{\rm{N_{CC_{Ar}}}}{\rm{N_{NC_{Ar}}}}$).
\item { {The ratio of} 
 $\nu_e$  {and} $\bar{\nu}e$  {events on Argon  in the DUNE detector}  {(}$\frac{\textrm{ {N}}{\nu_e-{Ar}}}{\textrm{ {N}}_{\bar{\nu}_e-{Ar}}}$ {)}}.
\end{enumerate}

By considering these combinations, the aim is to extract more information and enhance our sensitivity to neutrino mixing effects and the neutrino mass ordering in the supernova context.

These ratios are attractive because they do not depend on normalization factors (like the SN distance or the time-integrated luminosity $\epsilon$) and have already been used in the past to study the $\theta_{13}$ mixing angle and for disentangling between different neutrino flavor transformation scenarios \cite{Lunardini:2003,Capozzi:2018}. In this work, the extension of using these observables to study the neutrino mass ordering is explored. Additionally, variations in the parameters describing the initial neutrino spectrum within the ranges of 8 MeV $< \langle E_{\alpha} \rangle < $ 28 MeV { {and} $2 \leq \beta \leq 4.5$  {are examined}}. Furthermore, the constraint $\braket{E_{\nu_e}}\leq \braket{E_{\bar{\nu}_e}} \leq \braket{E{\nu_x,\bar{\nu}_x}}$ on the mean energies is imposed.

\begin{figure}[ht]
        \includegraphics[width=1.\textwidth]{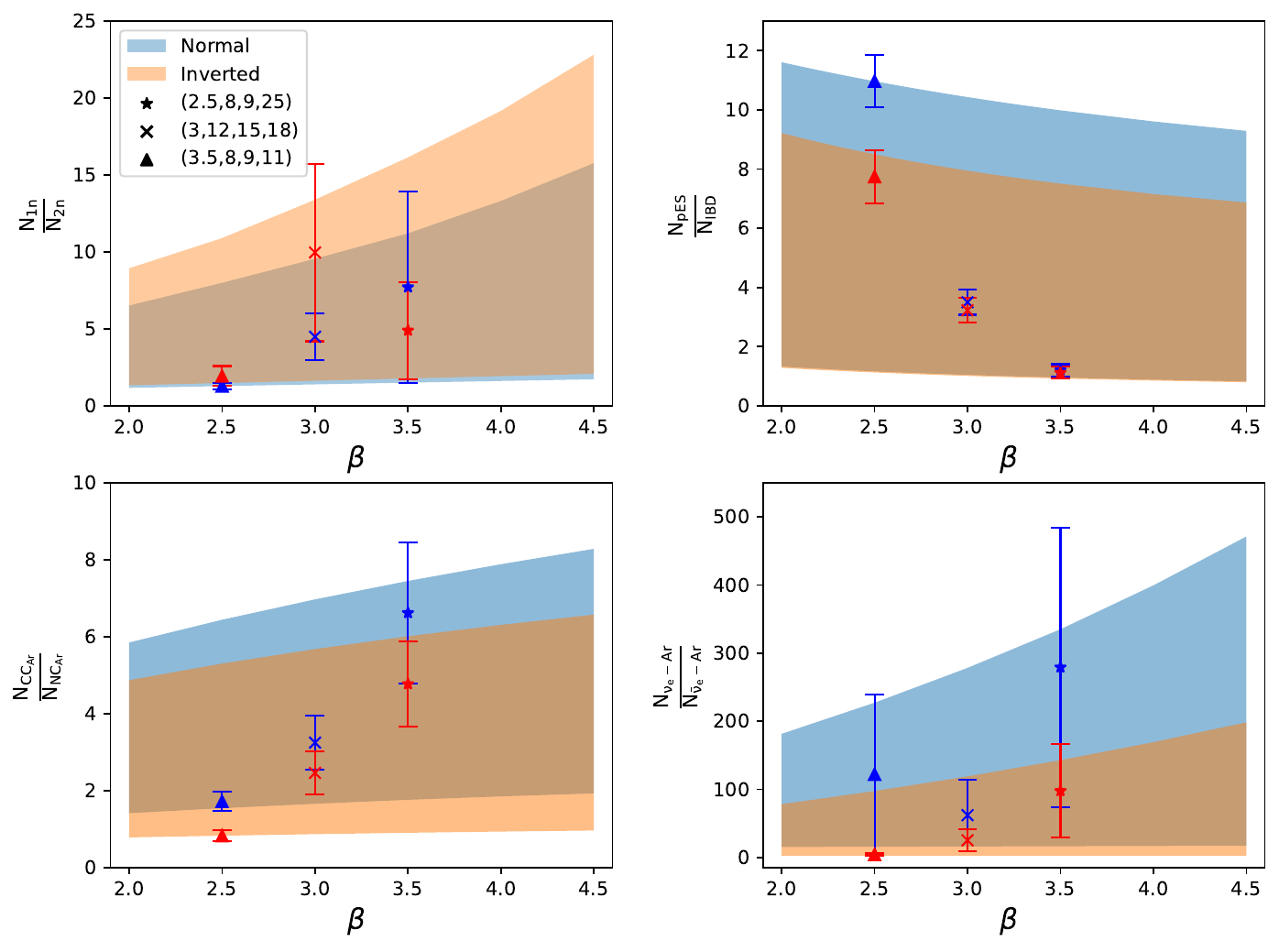}
        \caption{The four ratios studied as a function of the pinching parameter $\beta$.  Different colors indicate different mass ordering; the shaded regions are generated by the variations in the neutrino mean energies in the mentioned ranges. Different SN models were superimposed on the regions.  Triangles: Test Model 1; crosses: Dasgupta  {model} 
 \cite{Dasgupta:2011}; and stars: Test Model 2. The error bars show the relative statistical errors associated with the event ratios.
        } \label{all_maps}
    \end{figure}

In Figure~\ref{all_maps}, the ranges obtained for each ratio are displayed, considering both mass orderings and variations in the spectral parameters within the mentioned ranges.

Regarding the HALO detector ratio, it is observed that the $\nu_e +^{208}\textrm{Pb}$ reaction dominates, being prevalent in both 1n and 2n emission channels. Furthermore, a greater number of events are generated in the NO scenario compared to the IO scenario, and this difference is more pronounced in the 2n channel. Consequently, when analyzing the $\frac{\textrm{N}{1n}}{\textrm{N}{2n}}$ ratio, the IO scenario exhibits higher values than the NO case.

For the SNO+ ratio, IBD events in the IO scenario surpass those in the NO scenario. Consequently, the ratio $\frac{\textrm{N}{pES}}{\textrm{N}{IBD}}$ peaks in the NO scenario, as the $N_{pES}$ events remain unaffected by neutrino oscillations.

Shifting focus to the DUNE ratio $\frac{\textrm{N}_{CC{Ar}}}{\textrm{N}_{NC{Ar}}}$, the dominant channel for CC interactions is $\nu_e+^{40}\textrm{Ar}$. In this case, the NO scenario yields a higher event count than the IO scenario. This behavior is clearly depicted in the ratios presented in the bottom-left panel of Figure \ref{all_maps}.

In the case of the DUNE ratio $\frac{\textrm{N}{\nu_e-{Ar}}}{\textrm{N}{\bar{\nu}e-{Ar}}}$ (illustrated in the bottom-right panel of Figure \ref{all_maps}), this discrepancy is even more pronounced, given that $\textrm{N}_{\bar{\nu}e-{Ar}}$ is greater for the IO scenario compared to the NO one.

Regarding the behavior of the ratios with respect to the pinching parameter, it is observed that as $\beta$ increases, the individual counts in the studied channels $N_i$ decrease. This is related to the fact that smaller values of $\beta$ increase the number of neutrinos at higher energies and reduce the number of neutrinos at lower energies. 

For the ratios $\frac{\rm{N_{1n}}}{\rm{N_{2n}}}$, $\frac{\rm{N_{CC_{Ar}}}}{\rm{N_{NC_{Ar}}}}$, and $\frac{\textrm{N}{\nu_e-{Ar}}}{\textrm{N}{\bar{\nu}_e-{Ar}}}$, the decrease in counts is more pronounced for the denominator, resulting in higher values for the ratio as $\beta$ increases.

On the other hand, for the ratio $\frac{\rm{N_{pES}}}{\rm{N_{IBD}}}$, $\rm{N_{pES}}$ decreases more significantly compared to $\rm{N_{IBD}}$, leading to a lower ratio at higher values of $\beta$.

With experimental knowledge of the event ratios, determination of the allowed parameter ranges for each mass ordering would become feasible. In addition, by examining the overlaps between these ranges for the different studied ratios, it would be possible to determine the mass ordering.  Combining studies on different observables is crucial to avoid parameter degeneracies and obtain more robust results.

From  Figure \ref{all_maps}, it is seen that  $\frac{N_{1n}}{N_{2n}}>15.76$,  $\frac{N_{CC_{Ar}}}{N_{NC_{Ar}}}<1.41$, and  $\frac{\textrm{N}_{\nu_e-{Ar}}}{\textrm{N}_{\bar{\nu}_e-{Ar}}}<17.14$ exclude NO. Meanwhile, $\frac{N_{1n}}{N_{2n}}<1.33$, $\frac{N_{pES}}{N_{IBD}} > 9.21$, $\frac{N_{CC_{Ar}}}{N_{NC_{Ar}}}> 6.57$, and  $\frac{\textrm{N}_{\nu_e-{Ar}}}{\textrm{N}_{\bar{\nu}_e-{Ar}}}>198.29$ exclude IO. Additionally, calculations have been conducted to determine the values of these ratios for three distinct SN models. These are the Dasgupta model \cite{Dasgupta:2011} and two test models: \emph{Test Model 1} and \emph{Test Model 2}. Model 1 falls into non-overlapping regions for the HALO and SNO+ ratios and is given by $\beta=2.5$, $\braket{E_{\nu_e}}=8$, $\braket{E_{\bar{\nu_e}}} =9$, and $\braket{E_{\nu_x}}=25$, while the second falls in the non-overlapping region of the DUNE ratios and has a small dispersion for SNO+, and it is given by  $\beta=3.5$, $\braket{E_{\nu_e}}=8$, $\braket{E_{\bar{\nu_e}}} =9$, and $\braket{E_{\nu_x}}=11$. 
The three models have been superimposed on the figures along with their respective errors.

\begin{figure}[ht]
        \includegraphics[width=1.\textwidth]{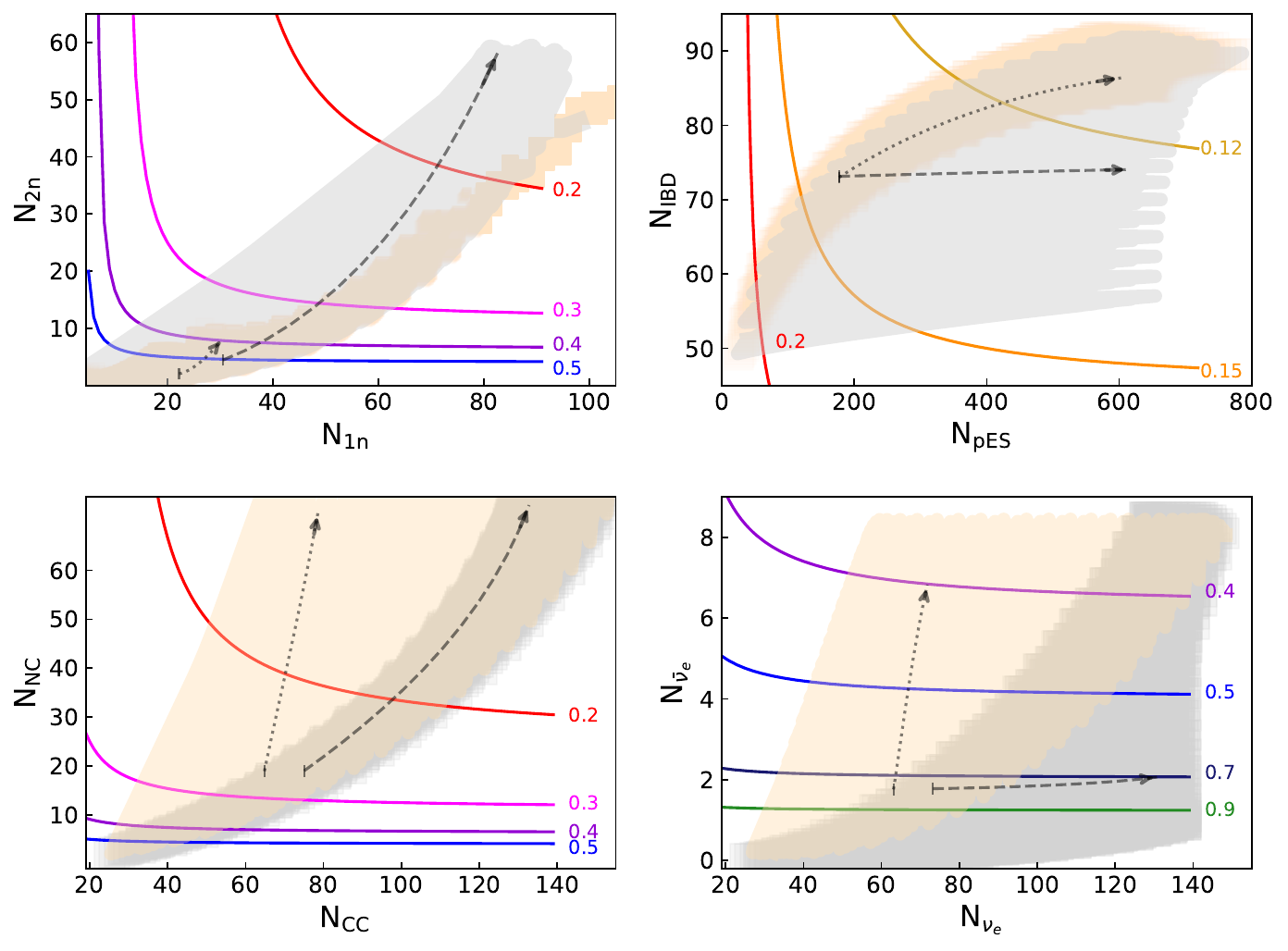}
        \caption{ Relative statistical error for the ratios shown in Figure \ref{all_maps} as a function of the numbers of events.  Above each contour line, the statistical error  is smaller than the value indicated in the figure. The shaded gray (beige) areas correspond to the counts that can be generated within the parameter space studied in this work considering NO (IO).}\label{error_maps}
    \end{figure}

\textls[-5]{In Figure \ref{error_maps}, the relative statistical error for the ratios depicted in Figure \ref{all_maps} is presented by solid lines\endnote{{The relative statistical} 
 error has been calculated as $\frac{\sigma(\frac{A}{B})}{\frac{A}{B}}=\sqrt{(\sigma_A/A)^2+(\sigma_B/B)^2}$.}. The shaded areas illustrate the counts that can be generated within the parameter space considered in this study. The gray or beige color corresponds to the counts generated by NO or IO, respectively. For this plot, the overall factors, d and $\epsilon$, were fixed at 10 kpc and 25 foe, respectively. In most of the studied cases, the statistics are sufficiently large, making the statistical error not a limiting factor. Furthermore, the arrows indicate the growth of the counts in each channel for NO (represented by dashed lines) and IO (represented by dotted lines). These arrows have been constructed with $\beta=3$, $\braket{E_{\nu_e}}=12 \, \rm{MeV}$, and $\braket{E_{\bar{\nu_e}}} =15 \, \rm{MeV}$, while the direction of the arrows indicates the increase in $\braket{E_{\nu_x}}$ from 15 MeV to 28 MeV.}

\section{Statistical Analysis}\label{stat}

{ {A statistical analysis was conducted to compare the expected event counts at each detector for both the NO and IO scenarios. The aim was to investigate the possibility of replicating an expected signal for a particular mass ordering using the opposite ordering while allowing for variations in the spectral SN parameters within a realistic range. The analysis employed a chi-square test to determine the feasibility of such a reproduction. Additionally, the study aimed to identify any preferential channels, detectors, or combinations thereof that could be instrumental in detecting signatures of the neutrino mass ordering.} 

\subsection{Individual Counts Analysis}

{ {First}
, }an assessment of the sensitivity of individual channel counts ($N_i$) within each detector was conducted using a chi-square minimization procedure. { {Subsequently},} the overall sensitivity was calculated by considering all the analyzed channels and detectors and minimizing a global chi-square. For the benchmark model, the NO scenario was considered. The indicator is defined by
\begin{equation}\label{eq:chisq_alpha}
    \chi^2_{\alpha}=\sum_i \frac{\left[N_i^\alpha(\hat{\theta}^{SN},NO)- N_i^\alpha(\theta^{SN},IO)\right]^2}{\sigma^2(N_i^\alpha(\hat{\theta}^{SN},NO))} \, ,
\end{equation}
hl{where} $\alpha$  {refers to the considered interaction channel}, $N_i^\alpha$  {are the number of events in the} $\alpha$  {channel,}  {calculated by integrating the spectra} $dN/dE$  {over the visible energy, as described in Section} \ref{detectors},  {and the sum runs over the} $i$- {energy bins}. For both SNO+ and DUNE, equally spaced bins of 0.5~MeV were considered \cite{Abi:2021,Stringer:2019ztj}. In the case of HALO, the analysis involved total counts because the detector operates based on neutron counting and does not entail a spectral analysis \cite{Volpe:2011}. For each data point, a statistical error of $\sigma_i=\sqrt{n_i}$ is considered.  $\hat{\theta}^{SN}$ and $\theta^{SN}$ represent the set of parameters $\{\beta,\braket{E_{\nu_e}},\braket{E_{\bar{\nu}e}},\braket{E{\nu_x}},\epsilon \}$ that define the initial SN spectra for the NO and IO models, respectively.  
Furthermore, the sensitivity is calculated by incorporating all the studied channels using a global chi-square, denoted as

\begin{equation}\label{eq:chisq_glob}
\chi^2_{\rm global}=\sum_\alpha \chi^2_{\alpha} \,.
\end{equation}

The exploration encompasses parameter ranges for both $\hat{\theta}^{SN}$ and $\theta^{SN}$, involving $8\, \textrm{MeV} \leq \braket{E_{\nu_j}} \leq 28 \,\textrm{MeV}$, $2<\beta<4.5$, and $1/6 \times 10^{53} \, \textrm{erg} < \epsilon < 1 \times 10^{53} \, \textrm{erg}$ (corresponding to $E_{\nu_{\text{tot}}}=[1-6]\times10^{53}$~erg). Additionally, the constraint $\braket{E_{\nu_e}} \leq \braket{E_{\bar{\nu}_e}} \leq \braket{E{\nu_x,\bar{\nu}_x}}$ is imposed on the average energies at the neutrinosphere.

The minimization of the statistical indicators was performed using  {IMINUIT}\footnote{IMINUIT v2.24.0 maintained by CERN's ROOT team \url{https://zenodo.org/records/8249703}} 
  \cite{iminuit}, a Python interface of the MINUIT2 C++ package (standard tool at CERN) that minimizes the multi-variate function with constraints \cite{James:1975dr}.

As a first step, we did not perform marginalization on $\epsilon$ but instead focused on 
studying the effects of its variation along with the SN distance. To accomplish this, we introduced an overall factor $A=\frac{\epsilon[\rm{foe}]}{d^2[\rm{kpc}]}$. It is important to note that the obtained results are applicable to any combination of $\epsilon$ and $d$ that yields the same value of A\endnote{{Taking the standard distance of }10 kpc, $A=0.01$, $A=0.25$, and $A=1$ correspond to luminosities of 1, 25, and 100 foe, respectively. Or, considering a fixed luminosity of 25 foe, the same A values correspond to distances of 5, 10, and 50 kpc, respectively.}. 

Concerning the analysis of the counts for the individual channels ($N_i$), it was observed that there was a lack of sensitivity to changes in the mass ordering. This was indicated by the fact that $\chi^2_{\alpha_{min}} <  0.6$ for all channels,  meaning that the counts produced by the NO could be reproduced using IO by adjusting the spectral parameters accordingly. In contrast, more significant sensitivities were identified when considering the global observable.
In Figure \ref{cont_global}, the results of the minimization process for the global chi-square are presented for various values of A. The contours illustrate the minimum chi-square as a function of the benchmark model parameters, taking into account the allowed variation in the IO model parameters as well. These provide insights into the parameter regions that enhance sensitivity.

It was observed that for the global observable, the sensitivity increased as the value of $\beta$ decreased and as the value of $\braket{E_{\nu_x}}$ increased. Qualitatively, this can be attributed to the extension of neutrino fluxes to higher energies in these conditions. Consequently, interaction channels with higher thresholds (such as the 2n channel for HALO and $\bar{\nu}_e-Ar$ for DUNE) become relevant and begin to substantially influence the fitting process. In turn, it was observed that the 1n and 2n channels in HALO, as well as the CCAr channels in DUNE, posed the most significant challenges for simultaneous minimization. As expected, higher values of A produce higher sensitivities, because they are associated either with shorter distances or higher luminosities.
 \begin{figure}[ht]
  \centering
        \includegraphics[width=1.0\textwidth]{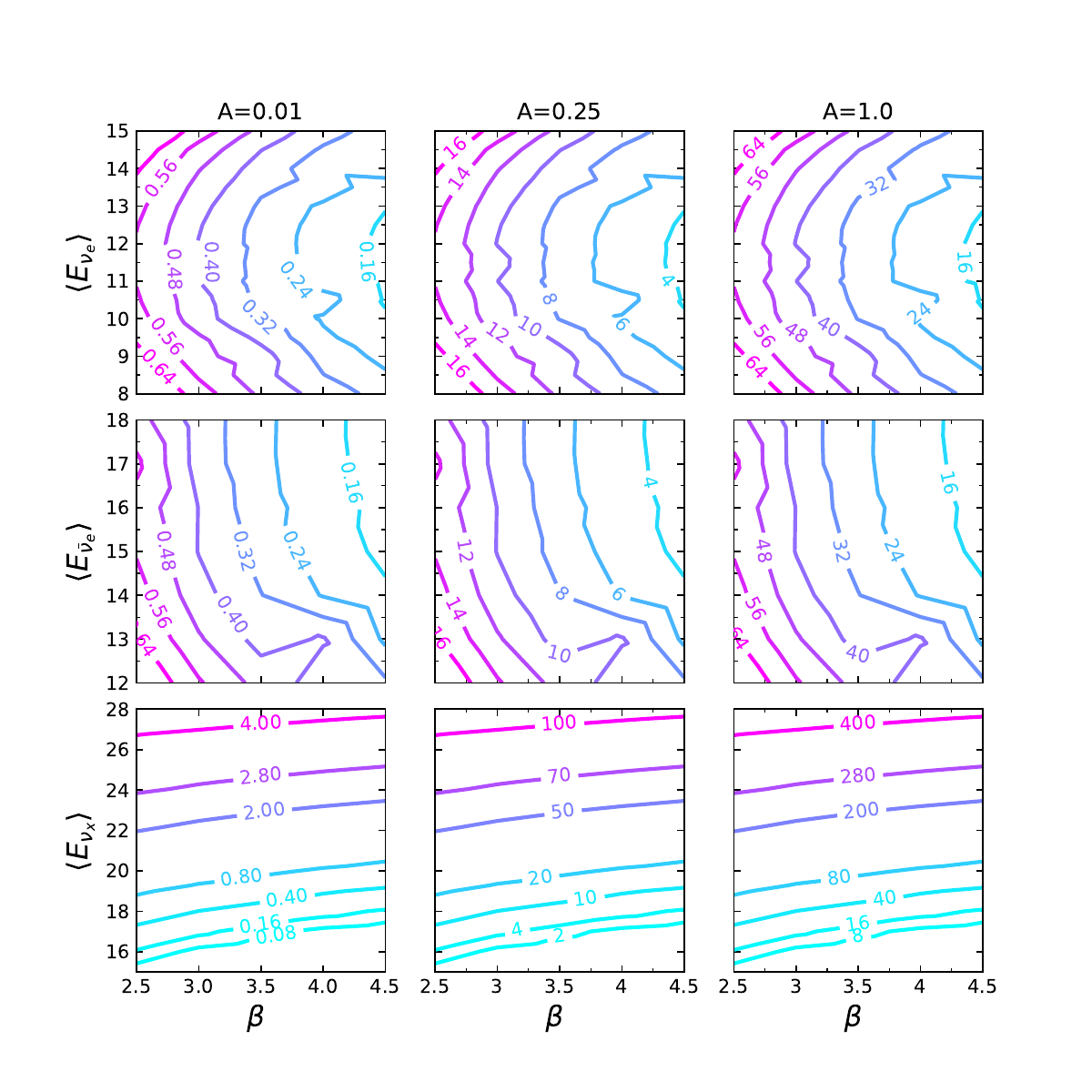} \caption{$\chi^2_{min}$  {contours} 
 as a function of the benchmark model parameters for the global observable  $ \chi^2_{global}=\sum_\alpha  \chi^2_{\alpha}$. The different columns stand for different values of A, as indicated.  For display purposes, in each row, the two remaining mean energies were fixed at values $\langle E_{{\nu}_e} \rangle =12$ MeV,{\linebreak} $\langle E_{\bar{\nu}_e} \rangle =15$ MeV, or $\langle E_{\nu_x} \rangle =18$ MeV. The different colors indicate different values of $\chi^2_{min}$, given by the numbers on the contours.}\label{cont_global}
    \end{figure}

Next, parameter A was incorporated into the minimization process. In Figure \ref{cont_global_epsilon}, contours for the global analysis are presented, considering the variation in the luminosity and distance between the models. It was observed that the minimization of the chi-square resulted in smaller values. This outcome was expected due to the incorporation of an additional marginalized parameter. A consistent observation was made that at higher non-electronic energies, there was a greater sensitivity in distinguishing between different models. Additionally, when considering higher luminosities or shorter distances, the discrimination between models with NO or IO becomes more feasible.

 \begin{figure}[H]
 
        \includegraphics[width=0.6\textwidth]{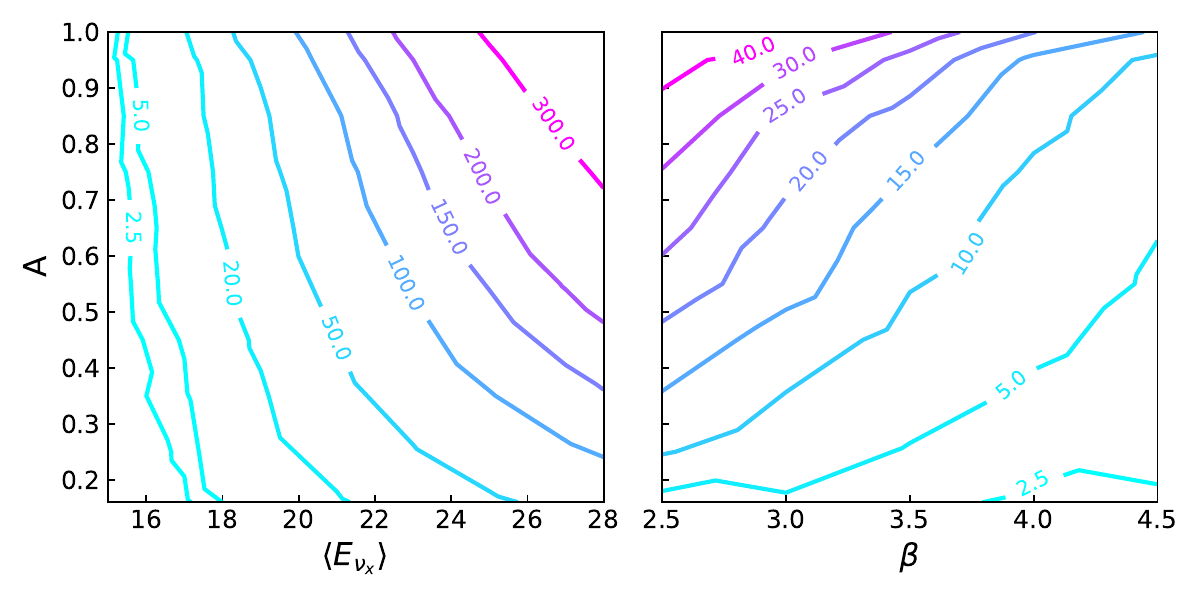} \caption{$\chi^2_{min}$  {contours} as a function of the benchmark model parameters $\beta$,  $\langle E_{\nu_x} \rangle$, and $A=\frac{\epsilon[\rm{foe}]}{d^2[\rm{kpc}]}$  for the global observable  $ \chi^2_{global}=\sum_\alpha  \chi^2_{\alpha}$.  For display purposes, the remaining parameters were fixed at values $\langle E_{{\nu}_e} \rangle =12$ MeV, $\langle E_{\bar{\nu}_e} \rangle =15$MeV, and $\beta=3$ for the left panel and at $\langle E_{{\nu}_e} \rangle =12$ MeV,{\linebreak} $\langle E_{\bar{\nu}_e} \rangle =15$ MeV, and $\langle E_{\nu_x} \rangle=18$ MeV for the right panel. The different colors indicate different values of $\chi^2_{min}$, given by the numbers on the contours.}\label{cont_global_epsilon}
    \end{figure}

\subsection{Ratio Analysis}
Next, a similar analysis was performed, focusing on the event ratios as defined in Section \ref{sec_ratios}. In this case: 
 \begin{equation}
    \chi^2_{\frac{ch_1}{ch_2}_{min}}=\frac{\left[\frac{N_{ch_1}}{N_{ch_2}}(\hat{\theta}^{SN},NO)- \frac{N_{ch_1}}{N_{ch_2}}(\theta^{SN},IO)\right]^2}{\sigma^2(\frac{N_{ch_1}}{N_{ch_2}}(\hat{\theta}^{SN},NO))}\, ,
\end{equation}
where $ch_1$ and $ch_2$ stand for the two involved channels in the ratio calculation. The errors were calculated as described in Section \ref{sec_ratios}. Unlike the previous case, the ratios  no longer depend on $\epsilon$ or distance. Thus, the results are not marginalized over epsilon. However, because epsilon contributes to the calculation of the errors, the results are presented for different values of A as defined in the previous subsection. The ranges for all the other parameters remain consistent with those mentioned above.
In Figure \ref{cont_ratios}, we show the contour plots of the minimum chi-square for the four studied ratios as a function of $\braket{E_{\nu_x}}$, because we noted that varying the values of mean energies for electron-type neutrinos has a small effect. Once again, we display the results for different values of the parameter A. 
Our findings indicate that there is an increased sensitivity with respect to the mass ordering as $E_{\nu_x}$ increases, with the exception of the $\frac{N_{CC_{Ar}}}{N_{NC_{Ar}}}$ ratio, which exhibits the opposite behavior. Additionally, we observed that the sensitivity of the ratios associated with HALO and SNO increases at lower values of $\beta$, while those associated with DUNE show enhanced sensitivity at higher values of $\beta$. 

The ratio associated with HALO exhibits the highest sensitivity. This can be attributed to the fact that, for large values of $E_{\nu_x}$ ($\gtrsim$18 MeV), the fluxes in the NO scenario extend to higher neutrino energies. Consequently, the $\nu_e +^{208}\rm{Pb} \rightarrow ^{206}\rm{Bi}+2n +e^{-}$ channel becomes significant for the NO ordering, resulting in larger event counts. As a result, there are substantial differences between the $\frac{\rm{N_{1n}}}{\rm{N_{2n}}}$  ratios for NO and IO, leading to an amplified chi-square numerator. Additionally, for these particular $E_{\nu_x}$ values, the event counts for the $\nu_e +^{208}\rm{Pb} \rightarrow ^{207}\rm{Bi}+n +e^{-}$ channel increase for both mass orderings. The combined effect of these factors results in a large number of counts on both channels 1n and 2n, which are associated with smaller statistical errors, reducing the denominator of the chi-square (as depicted in Figure \ref{error_maps}) and thereby enhancing the overall sensitivity. 

The second dominant ratio associated with SNO+ ($\frac{\rm{N_{pES}}}{\rm{N_{IBD}}}$) exhibits significant growth at high $\braket{E_{\nu_x}}$. This growth can be primarily attributed to a substantial increase in counts from both the pES channel and the IBD channel for IO, indicating a strong response at non-electronic energies. In contrast, the counts in the IBD channel for the NO scenario remain relatively unchanged. Consequently, as $\braket{E_{\nu_x}}$ increases, the numerator of the corresponding chi-square becomes larger. Moreover, the increase in counts in the pES channel leads to a smaller sigma error, ultimately resulting in a higher final chi-square value. 

 \begin{figure}[H]
 
        \includegraphics[width=1.0\textwidth]{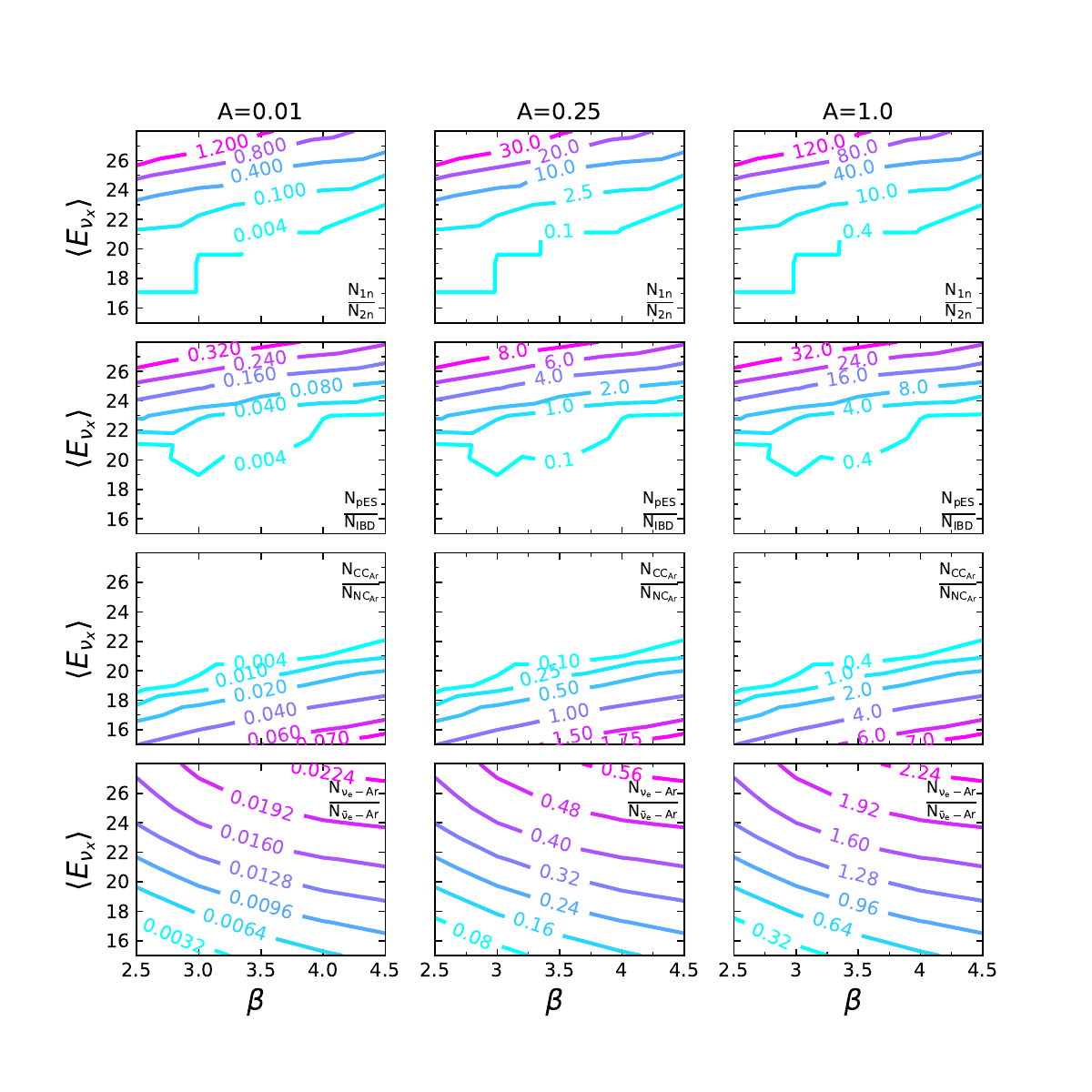} \caption{$\chi^2_{min}$ contours as a function of the benchmark model parameters $E_{\nu_x}$ and $\beta$ for the proposed ratios of events. Different rows stand for different observables, as indicated in the figure. The different columns stand for different values of A, as indicated.  For all the plots,  $\langle E_{{\nu}_e} \rangle =12$ MeV and $\langle E_{\bar{\nu}_e} \rangle =15$ MeV. The different colors indicate different values of $\chi^2_{min}$, given by the numbers on the contours.}\label{cont_ratios}
    \end{figure}

\section{Conclusions}
The neutrino mass ordering problem remains unsolved;  thus, it is interesting to investigate if a future observation of a galactic SN could help to discriminate among possible mass ordering scenarios.

In this work, we have presented predictions for the expected neutrino events in three different detectors with distinct technologies: the scintillator SNO+, the helium and lead detector HALO, and the liquid argon detector DUNE. 

First, we presented a study regarding the ratios of events, which let us eliminate the dependence with the distance and with the integrated luminosity.  This allowed us to focus specifically on how these ratios vary with respect to the SN model spectral parameters and the mass ordering.

We showed that the ratios $\frac{N_{1n}}{N_{2n}}$, $\frac{N_{pES}}{N_{IBD}}$,  $\frac{N_{CC_{Ar}}}{N_{NC_{Ar}}}$, and  $\frac{\textrm{N}_{\nu_e-{Ar}}}{\textrm{N}_{\bar{\nu}_e-{Ar}}}$  can potentially exclude either normal or inverted mass ordering for the neutrino sector without requiring the assumption of a specific SN model.

We were able to find regions that exclude one ordering or another, finding the following:
\begin{itemize}
    \item $\frac{N_{1n}}{N_{2n}}>15.76$  excludes NO, while $\frac{N_{1n}}{N_{2n}}<1.33$ excludes IO;
    \item $\frac{N_{pES}}{N_{IBD}} > 9.21$ excludes IO;
    \item $\frac{N_{CC_{Ar}}}{N_{NC_{Ar}}}<1.41$ excludes NO, while  $\frac{N_{CC_{Ar}}}{N_{NC_{Ar}}}> 6.57$ excludes IO; 
    \item $\frac{\textrm{N}_{\nu_e-{Ar}}}{\textrm{N}_{\bar{\nu}_e-{Ar}}}<17.14$, excludes NO, while $\frac{\textrm{N}_{\nu_e-{Ar}}}{\textrm{N}_{\bar{\nu}_e-{Ar}}}>198.29$ excludes IO. 
\end{itemize}

Then, we performed a statistical analysis to study the sensitivities of different observables associated with the change in the neutrino mass ordering. We had observed that analyzing combinations of events across different channels yielded more meaningful insights than examining the isolated counts for individual channels ($N_i$). In particular, we observed that the global observable, encompassing all channels collectively, displayed a heightened sensitivity in discerning between the mixing scenarios. Specifically, the 1n and 2n channels in HALO, along with the CCAr channels in DUNE, posed significant challenges during simultaneous minimization, identifying them as crucial channels.

Through statistical analysis and minimization, we found that the most sensitive ratios to the mass ordering are 
 $\frac{\rm{N_{\rm{1n}}}}{\rm{N_{\rm{2n}}}}$ for HALO and $\frac{\rm{N}_{pES}}{\rm{N}_{IBD}}$ for SNO+, being two complementary detectors when studying these effects. The fact that they have the same location makes this even more interesting because there will be no effects on the signal due to geographical differences, such as the Earth matter effects. Additionally, the ratio $\frac{\textrm{N}_{\nu_e-{Ar}}}{\textrm{N}_{\bar{\nu}_e-{Ar}}}$ for DUNE is the one that generates larger numerical differences when comparing both mass orderings, but it is also the one with the greatest associated statistical errors. 

Although the specific parameters governing future supernova explosions remain unknown, exploring the parameter space within the ranges studied reveals a wide range of models that could exhibit high sensitivity. By examining the sensitivities of various observables, we can determine which combination of experiments would be sensitive in different regions of the parameter space.
Through the analysis of multiple signals, considering different channels and ratios, we have the potential to differentiate between different mass orderings. The probability of achieving higher sensitivities for this discrimination increases if future observations occur in the regions of Figure \ref{all_maps} that do not overlap.
In general, we observe that higher non-electronic energies offer greater sensitivity in distinguishing between different models, while higher luminosities or shorter distances enhance the discrimination between models with NO or IO, making it more feasible to identify the ordering of neutrino masses.

 The data collected from a network of detectors with diverse energy thresholds and detection channels offers a unique and powerful tool for imposing constraints on the parameters of the initial SN neutrino fluxes and determining the neutrino mass ordering. This approach provides valuable insights into the influence of SN spectral parameters and mass ordering on the observed data, leading to a more comprehensive understanding of the underlying physics involved in these phenomena.

\vspace{10pt}
\funding{Short term grant by German Academic Exchange Service (DAAD)}

\dataavailability{The data presented in this study are available on request from the corresponding author.} 

\acknowledgments{ {I extend my gratitude to the German Academic Exchange Service (DAAD) for awarding me a short-term grant, enabling a 3-month research stay at the Theoretical Astroparticle Physics group, led by Thomas Schewtz-Mangold, at the Karlsruhe Institute of Technology (KIT). Their generous support has played a pivotal role in the advancement of this work. I appreciate the fruitful discussions and guidance given by Belina von Krosigk and Thomas Schewtz-Mangold; their contribution has greatly enriched this work. I acknowledge the Interdisciplinary Theoretical and Mathematical Sciences Program (iTHEMS) at RIKEN and the Network for Neutrinos, Nuclear Astrophysics, and Symmetries (N3AS) for providing spaces for enriching discussions during the development of this research.} 
}

\conflictsofinterest{The author declares no conflicts of interest.} 

\appendixtitles{yes} 
\appendixstart
\appendix
\section[\appendixname~\thesection]{Neutrino Flavor Evolution Equations} \label{evol_eq}
To solve Equation (\ref{flavor-evol}), it is also necessary to recalculate in a coupled manner the reaction rates, the baryon density, and the electronic fraction of the material, because all these quantities will be sensitive to the effects of the oscillations. The neutrino scattering rates are functions of the neutrino fluxes and then the flavor oscillations cannot be neglected. Also,  the weak reactions will modify the amount of neutrons, protons, and electrons in the star through  neutrino- and antineutrino-induced reactions
\begin{eqnarray}
\label{nu-n}
\nu_e+n &\rightarrow &p +e^- \, , \nonumber\\
\label{nu-p}
\bar{\nu}_e+p &\rightarrow& n + e^+ \, .
\end{eqnarray}

The rate of these two reactions can be computed as \cite{Balantekin:2004}
\begin{eqnarray}
\lambda_\nu&=&\int \sigma_\nu(E_\nu)\frac{d\phi_\nu}{dE_\nu }dE_\nu \, ,
\end{eqnarray}
where the cross sections, in units of ${\rm cm}^2$, are
\begin{eqnarray}
\sigma_{\nu_e}(E_{\nu_e})&=&9.6\times10^{-44}\left(\frac{E_{\nu_e}+\Delta m_{np}}{MeV}\right) \, , \\
\sigma_{\bar{\nu}_e}(E_{\bar{\nu}_e})&=&9.6\times10^{-44}\left(\frac{E_{\bar{\nu}_e}-\Delta m_{np}}{MeV}\right)\, .
\end{eqnarray}

In the last expressions, $\Delta m_{np}= 1.293 \, \rm{MeV}$ is the neutron-to-proton mass difference. Notice that for the antineutrino cross section $\sigma_{\bar{\nu}_e}$, the antineutrino energy must be larger than the neutron-to-proton mass difference $\left(E_{\bar{\nu}_e}>\Delta m_{np}\right)$.

The reaction rates for the inverse reactions of Equation (\ref{nu-n})  are written \cite{Tamborra:2012}: 
\begin{eqnarray} \label{inv_velos}
\lambda_{e^-} & \simeq & 1.578\times 10^{-2}\left(\frac{T_e}{m_e}\right)^5 e^{(-1.293+\mu_e)/T_e}
\left(1+\frac{0.646~\rm{MeV}}{T_e}+\frac{0.128~\rm{MeV^2}}{T_e^2} \right) \, ,\nonumber\\
\lambda_{e^+} & \simeq & 1.578\times 10^{-2}\left(\frac{T_e}{m_e}\right)^5 e^{(-0.511-\mu_e)/T_e}\nonumber \\&&\times
\left(1+\frac{0.1.16~\rm{MeV}}{T_e}+\frac{0.601~\rm{MeV^2}}{T_e^2}
+\frac{0.178~\rm{MeV^3}}{T_e^3}+\frac{0.035~\rm{MeV^4}}{T_e^4} \right) \, . 
\end{eqnarray}

In these expressions, $m_e$, $\mu_e$, and $T_e$ are the electron mass, the electron chemical potential, and the electron temperatures (in units of ${\rm MeV}$), and the rates are given in units of ${\rm s}^{-1}$.
The electron chemical potential, at a fixed temperature, can be obtained following \mbox{reference \cite{Tamborra:2012}.}

If the environment is electrically neutral, the electron fraction can be computed in terms of the weak reaction velocities of Equations (\ref{nu-n}) and (\ref{inv_velos}). 
In the absence of heavy elements, the time dependence of the electron fraction is equal to the one of protons. Also, in this case, the sum of the fraction masses of neutrons, protons, and $\alpha$-particles is given by the relation $X_p+X_n+X_\alpha=1$, giving:
\begin{eqnarray}
\frac{dY_e}{dt}&=&\lambda_n-(\lambda_p+\lambda_n)Y_e+\frac{1}{2}(\lambda_p-\lambda_n)X_\alpha \, ,
\end{eqnarray}
where $\lambda_p=\lambda_{\bar{\nu}_e}+\lambda_{e^-}$ and $\lambda_n=\lambda_{\nu_e}+\lambda_{e^+}$.

If the plasma reaches a stage of weak equilibrium, the electron fraction does not change with time, that is, $\frac{dY_e}{dt}=0$, and therefore
\begin{eqnarray} \label{eq:Ye}
Y_e&=&\frac{\lambda_n}{\lambda_n+\lambda_p}+\frac{1}{2}\frac{\lambda_p-\lambda_n}{\lambda_p+\lambda_n} X_\alpha \, ,
\end{eqnarray}
where 
 $X_{\alpha}$ is the fraction of the alpha particles.

For the mass Hamiltonian in vacuum from eq. \ref{flavor-evol}, we consider:
\begin{equation}
H^{vac}_{mass}=\begin{pmatrix} pc+ \frac{m_1^2c^4}{2E}& 0 \\ 0 & pc+ \frac{m_3^2c^4}{2E}  \end{pmatrix}
\end{equation}
 where  $m_i$ is the mass of the mass eigenstate $i$, $p$ stands for the momentum, and $c$ is the speed of light in a vacuum.
Considering the mixing matrix,
\begin{equation}
U=\begin{pmatrix}c_{13} & s_{13} \\ -s_{13} & c_{13} \end{pmatrix}
\end{equation} in which we have called $s_{ij}=\sin(\theta_{ij})$ y $c_{ij}=\cos(\theta_{ij})$, 
we can transform to the flavor base~by
\begin{equation}
H^{vac}_E=UH^{vac}_{mass}U^{\dagger}
\end{equation}
obtaining 
\begin{equation}
H^{vac}=\left(pc+\frac{m_1^2c^4}{2E}\right)\begin{pmatrix}1 & 0\\ 0 & 1\end{pmatrix}+\frac{\Delta_{13}c^4}{2E}\begin{pmatrix} s^2_{13}&\bar{s}_{13}\\ \bar{s}_{13}& c^2_{13}\end{pmatrix}
\end{equation}
where $\bar{s}_{ij}=\frac{\sin(2\theta_{ij})}{2}$ and $\Delta_{13}=m^2_3-m^2_1$.

Then, if the neutrinos travel through material media, they undergo coherent elastic scattering with other particles. Neutrino--neutron and neutrino--electron interactions are described by the MSW Hamiltonian \cite{Wolfenstein:1977}

\begin{equation}
H^m=\sqrt{2}G_F\begin{pmatrix}N_e-\frac{N_n}{2}&0\\0&-\frac{N_n}{2}\end{pmatrix}
\end{equation}
where $N_e$ and $N_n$ are the electron and neutron densities, respectively. If we consider an electrically neutral medium and neglect the presence of heavy particles, the above Hamiltonian can be written in terms of $Y_e$ given by Equation (\ref{eq:Ye}) reading
\begin{equation}
H^{m}=\frac{\sqrt{2}}{2}G_FN_b\begin{pmatrix}3Y_e-1&0\\0&Y_e-1\end{pmatrix}
\end{equation}
where  $G_F$ is the Fermi constant and $N_b$ is the baryon density \cite{Balantekin:2004}.

\section{Detection Channels in SNO+}\label{channels}

\paragraph{ {Inverse beta decay (IBD):}} 

Neutrinos can interact with free protons through inverse beta decay
$\left(\bar{\nu}_e + p \rightarrow n + e^+\right)$: 
the expected number of events can be computed as \cite{Strumia:2003}
\begin{equation}
\label{num-eventos}
N=N_p\int_{E_{min}}^\infty dE F_{\bar{\nu}_e} (E) \sigma_{\bar{\nu}_{e}}(E) \, \, ,
\end{equation}
where $E_{min}= 1.806 \, \rm{MeV}$ is the IBD energy threshold and $\sigma_{\bar{\nu}_{ep}}$ its cross section which, in units of $10^{-43}\, {\rm cm}^2$, can be approximated as \cite{Strumia:2003}
\begin{equation}\label{sig(e)}
\sigma_{\bar{\nu}_{e}}(E)=p_e E_e E^{-\,0.07056\, +\,0.02018\, \ln E\,-\,0.001953 \, \ln E^3} \, \, ,
\end{equation}
where $p_e$ is the positron momentum related to the neutrino energy $E$ as $p_e^2=(E-\Delta)^2-m_e^2$, the neutron to proton mass difference is $\Delta=m_n-m_p=1.293 \, {\rm MeV}$, and $m_e$ is the positron mass. 
The neutrino and positron energies are related by $E_e=E -\Delta$.
By assuming a $3\,{\rm kT}$ detector of ${\rm C}_6 {\rm H}_5 {\rm C}_{12} {\rm H}_{25}$ \cite{SNOplus:2015}, the number of free protons is $\rm{N}_p=2.2\times 10^{32}$.

\paragraph{Neutrino--proton elastic scattering (pES):}

The reaction $\nu +p \rightarrow \nu'+p$ is possible for all flavors. Although the cross section of the process is three times smaller than the IBD one, it is the channel that reports the greatest number of events because of the contribution of all six flavor neutrinos. 
The complete cross section for this process can be written as \cite{Weinberg:1972,Ahrens:1987}:
\begin{adjustwidth}{-\extralength}{0cm}
\begin{equation}
    \frac{d\sigma}{dE_p}(E_\nu,E_p)=\frac{G_F^2 m_p}{2\pi E_\nu^2}\left[(C_V\pm C_A)^2 E_\nu^2+(C_V \mp C_A)(E_\nu-E_p)^2 -(C_V^2-C_A^2)m_pE_p\right]\,,
\end{equation}
\end{adjustwidth}
where $E_p$ is the proton recoil energy and $E_\nu$ the incoming neutrino one. The upper (lower) sign corresponds to neutrinos (antineutrinos). 
$C_V=1/2-2\sin^2\theta_w$ and $C_A=\frac{g_A(0)(1+\eta)}{2}$ are the vector and axialvector coupling constants, respectively; $\theta_w$ is the effective weak mixing angle ($\sin^2 \theta_w = 0.23155$); $g_A(0)\sim 1.26$ is the axial proton form factor~\cite{Beringer:2012}; and $\eta$ is the proton strangeness, which is the contribution of the s-quark to $g_A(0)${\linebreak} ($\eta=0.12\pm0.07$) \cite{Ahrens:1987}. 
The true proton recoil spectrum can be calculated as
\begin{equation}\label{rate-pES}
    \frac{dN}{dE_p}(E_p)=N_p\int^\infty_{E_\nu^{min}}\frac{d\sigma}{dE_p}(E_\nu) F_\nu dE_\nu 
\end{equation}
with $N_p=3.32\times10^{31}$ protons in a fiducial mass of 0.45kt and ${E_\nu^{min}}=\frac{E_p+\sqrt{E_p(E_p+2m_p)}}{2}$ is the minimum neutrino energy required to reach a distinct $E_p$.
The visible energy $E_{vis}$, {i.e.,} the energy measured by the detector, is strongly quenched with respect to $E_p$. 
To convert the true proton energy spectra into the spectra of the visible electron energy $E_{vis}$, we applied Equation (\ref{rate-pES}) to a proton quenching factor $Q_p(E_p)$ extracted from Ref. \cite{vonkrosigk:2013}, which follows a Birk's law with $kB = 0.0098$ cm MeV$^{-1}$ and C = 0 m$^2$MeV$^{-2}$.

\paragraph{Neutrino$-$nucleus reactions in $^{12}$C:}

SN neutrinos can interact through the neutral current process $\nu + {^{12}C} \rightarrow {^{12}C}*(15.11\textrm{MeV}) +\nu'$, whose distinctive feature is the emission of a $15.11$~MeV $\gamma$ cascade due to the de-exitation of the scattered nucleus that produces a distinct spike in the energy spectrum. 
For this NC reaction, only the isovector axial current contributes to the interaction, and the cross section is given by \cite{Armbruster:1998}:
\begin{equation}
\sigma(E_\nu)=1.08\times 10^{-38}cm^2 \left(\frac{E_\nu-w}{M_N}\right)^2 \beta^2 k^2\,,
\end{equation}
where $E_\nu$ is the energy of the incident neutrino, $w$ is the negative Q$-$value of the reaction ($-15.11~$MeV), $M_N$ is the nucleon mass and $\beta$ is the isovector$-$axialvector coupling. 
The $k$ constant expresses the relative strength of the neutral and charged currents. 
Here, we have taken  $k=\beta=1$ according to the Standard Model (SM) \cite{Armbruster:1998,Donnelly:1979}.

\section{Some Comments about Collective Effects}\label{coll}  
 {In the deepest} regions inside an SN, neutrinos frequently undergo forward-scattering interactions with other oscillating neutrinos, giving rise to intriguing collective flavor oscillations. As mentioned earlier, during early times such as the neutronization burst and accretion phase, it is expected that the matter potential dominates over the neutrino--neutrino potential, strongly suppressing self-induced effects, in particular, those associated with the so-called \emph{slow} transformations\endnote{{The collective flavor oscillations are governed} by the neutrino--neutrino forward-scattering rate, $\mu=\sqrt{2}G_Fn_\nu$. The rate at which these oscillations occur is referred to as  {\emph{slow}} 
 if it is on the order of $\sim \sqrt{\mu\langle\omega\rangle}$ or  {\emph{fast}} if it is on the order of $\mu$, where $\langle\omega\rangle$ represents the collective synchronized rate.} that can induce spectral swaps and splits \cite{Chakraborty:2011,Sarikas:2011}. In such cases, the flavor evolution of neutrinos is determined solely by the influence of matter, as assumed in the previous sections.

However, the aforementioned suppression has recently been challenged \cite{Bhattacharyya:2022,Richers:2021,Wu:2021,Dasgupta:2015}. Several studies suggest that the presence of temporal instabilities in the dense neutrino gas allows for self-induced effects, even in the presence of a dominant matter density \cite{Dasgupta:2015}. Furthermore, fast flavor conversions occurring just above the supernova core may not be inhibited by high matter density \cite{Bhattacharyya:2022,Sawyer:2016,Dasgupta:2017}, potentially leading to flavor decoherence and the equalization of fluxes and spectra among different neutrino species \cite{Capozzi:2018,Sawyer:2016}. If fast flavor conversions or temporal instabilities occur in the deepest regions of the supernova, unimpeded by matter effects, they could tend to equalize the different fluences.
Despite attempts to characterize these effects, our current understanding of them remains far from settled.

Including these effects in the calculations we have performed in this study is not a simple task. However, a good step in the direction of understanding the eventual outcome of flavor transformation in supernovae has been taken in the reference \cite{Bhattacharyya:2022}, where some suggestions for supernova modelers are provided.
To implement these guidelines, a precise understanding of the angular distribution of the initial SN  model is needed. Regrettably, for this work, this remains unattainable due to the absence of a specific simulation under consideration. Consequently, we are constrained to adopt a naive approximation, achieved by equalizing the fluxes.

\begin{equation}
    F_{\nu_e}= F_{\nu_\mu}=F_{\nu_\tau}= \frac{F^0_{\nu_e}+ F^0_{\nu_\mu}+F^0_{\nu_\tau}}{3}
\end{equation}
In the presence of partial flavor equalization (FE), the total survival probability would be intermediate between the one given by only the matter effect ($\sim 0$ for NO and $\sim 0.3$ for IO) and the one obtained for complete flavor equalization ($1/3$).

Due to the inherent challenges associated with the intricate modeling of these effects, we only studied the potential discernment between the ME and a scenario primarily influenced by matter effects (NO or IO) within the initial one-second signal. This differentiation is pursued through the utilization of the ratios as defined in Section \ref{sec_ratios}. The notion of distinguishing between these scenarios was also explored in a prior work  \cite{Capozzi:2018}, in which an analysis was conducted of the IBD and pES channels at HyperK and the ArCC channel at DUNE. We examined the following indicator: 

 \begin{equation}
    \chi^2_{\frac{ch_1}{ch_2}}=\frac{\left[\frac{N_{ch_1}}{N_{ch_2}}(\hat{\theta}^{SN},FE)- \frac{N_{ch_1}}{N_{ch_2}}(\theta^{SN}, NO \,or\, IO)\right]^2}{\sigma^2(\frac{N_{ch_1}}{N_{ch_2}}(\hat{\theta}^{SN},FE))}\, ,
\end{equation}
where  $\theta^{SN}$ represent the set of parameters $\{\beta,\braket{E_{\nu_e}},\braket{E_{\bar{\nu}e}},\braket{E{\nu_x}} \}$ that define the initial SN spectra for the FE and the NO and IO models.  
In this scenario, a comparison is made between models with identical parameters. A more comprehensive analysis is planned for the future. Once again, the outcomes demonstrate sensitivity to $\braket{E_{\nu_x}}$. Figure \ref{chi_FE} presents the $\chi^2$ behavior with respect to energy variation. Various colors denote the distinct ratios under investigation, while the dashed or dotted lines indicate the comparison of FE with NO or IO, respectively. Notably, when comparing FE with NO, the ratio $\frac{\textrm{N}_{\nu_e-{Ar}}}{\textrm{N}_{\bar{\nu}e-{Ar}}}$ proves to be the most sensitive, followed by the ratio $\frac{\textrm{N}_{pES}}{\textrm{N}_{IBD}}$. Conversely, in comparison with the IO case, the two most sensitive ratios are $\frac{\textrm{N}_{1n}}{\textrm{N}_{2n}}$ and $\frac{\textrm{N}_{CC_{Ar}}}{\textrm{N}_{NC{Ar}}}$.
As in the previous sections, these sensitivities can be explained by  taking into account the behavior of the fluxes, as well as the cross sections and thresholds of each channel. Qualitatively, we have found that the counts associated with the $\bar{\nu}_e + ^{40}Ar$ channel for the FE case increase considerably with respect to those produced by matter effects, because the tail of the $\bar{\nu}_e$ distribution rises with respect to the case with NO, reaching the $\sim 18\, $MeV  threshold more easily (see \mbox{Figure \ref{total-Ar}}). On the other hand, the case with FE produces more counts in both channels than IO, but the ratio between the two remains approximately the same. This produces the trend $\frac{\textrm{N}_{\nu_e-{Ar}}} {\textrm{N}_{\bar{\nu}_e-{Ar}}}(FE) \approx \frac{\textrm{N}_{\nu_e-{Ar}}} {\textrm{N}_{\bar{\nu}_e-{Ar}}} (IO) < \frac{\textrm{N}_{\nu_e-{Ar}}} {\textrm{N}_{\bar{\nu}_e-{Ar}}} (NO)$. 

In the case of the IBD reaction, the threshold of the reaction is low ($\sim$ 1.806 MeV), which means that the behavior of the entire distribution is relevant, not just the tail. Because of this, FE produces slightly more counts than the NO case and practically the same amount as the IO case. As expected, the counts remain unchanged for the pES channel. This generates $\frac{\rm{N_{pES}}}{\rm{N_{IBD}}} (FE) \approx \frac{\rm{N_{pES}}}{\rm{N_{IBD}}}(IO) < \frac{\rm{N_{pES}}}{\rm{N_{IBD}}} (NO)$. 

For the HALO ratio, the counts produced in the 1n and 2n channels are similar for the FE and NO cases. In the case with IO, the tail in the distribution of $\nu_e$ is lower than that obtained with FE, generating fewer counts, especially in the $\nu_e +^{208}\rm{Pb} \rightarrow ^{206}\rm{Bi}+2n +e^{-}$  channel (due to its high threshold). In this case, we found that 
$\frac{\rm{N_{1n}}}{\rm{N_{2n}}}(FE) \approx \frac{\rm{N_{1n}}}{\rm{N_{2n}}} (NO) < \frac{\rm{N_{1n}}}{\rm{N_{2n}}}(IO)$. 

Finally, for $\frac{\rm{N_{CC_{Ar}}}}{\rm{N_{NC_{Ar}}}}$ at DUNE, the counts associated with NC remain constant, while those of the CC channels produce lower counts for IO than for NO or FE, as mentioned before. 
Thus,  $\nu_e+^{40}\rm{Ar} \rightarrow e^- + ^{40}\rm{K}^*$, the dominant channel. In this case, the trend is $\frac{\rm{N_{CC_{Ar}}}}{\rm{N_{NC_{Ar}}}} (IO) < \frac{\rm{N_{CC_{Ar}}}}{\rm{N_{NC_{Ar}}}} (NO) \approx \frac{\rm{N_{CC_{Ar}}}}{\rm{N_{NC_{Ar}}}} (FE) $.

The ratios $\frac{\textrm{N}_{\nu_e-{Ar}}} {\textrm{N}_{\bar{\nu}_e-{Ar}}}$ and  $\frac{\rm{N_{1n}}}{\rm{N_{2n}}}$ are the most affected by the change in $\braket{E_{\nu_x}}$, because their sensitivities increase as a consequence of the modification in the tail of the distributions and the thresholds of the $\bar{\nu}_e+^{40}\rm{Ar} \rightarrow e^+ + ^{40}\rm{Cl}^*$ and $\nu_e +^{208}\rm{Pb} \rightarrow ^{206}\rm{Bi}+2n +e^{-}$ reactions, respectively. 

The identified observables present valuable potential for distinguishing between the two mixing scenarios. However, a more realistic modeling of the collective oscillations is required to provide precise values and predictions.  A comprehensive study of this nature is reserved for future research.
\vspace{-10pt}
\begin{figure}[H]
\includegraphics[width=.7\textwidth ]{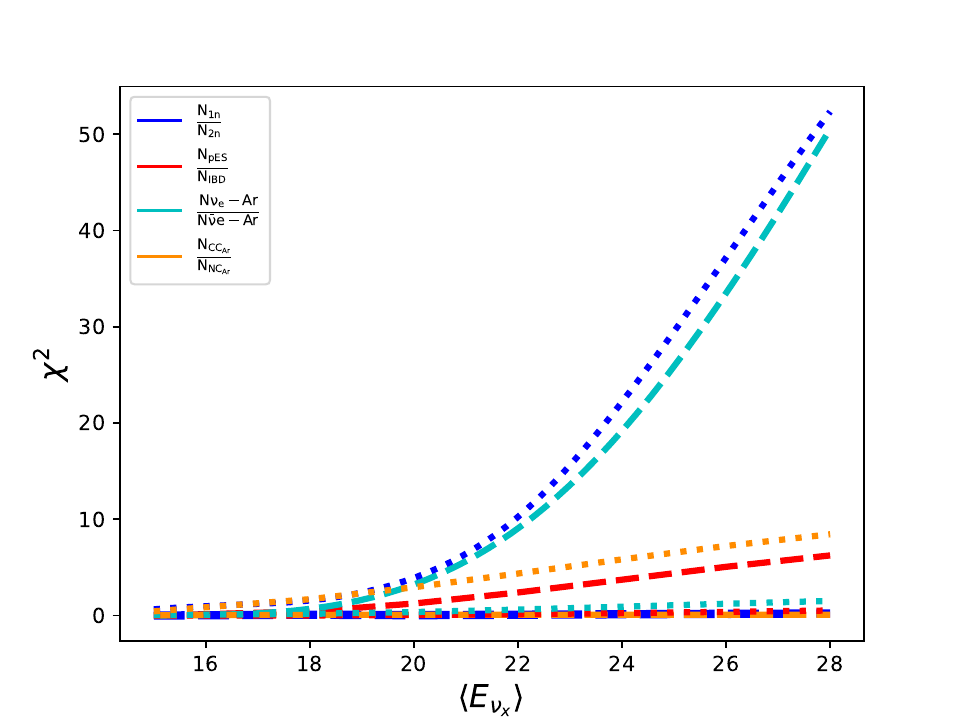}
\caption{Chi-square as a function of $\braket{E_{\nu_x}}$; the rest of the parameters remain fixed at   $\beta=3$, $\braket{E_{\nu_e}}=12$ MeV, and $\braket{E_{\bar{\nu}_e }}=15$ MeV. Dashed lines: comparison with NO; dotted lines: comparison with IO.}  
\label{chi_FE}
\end{figure}

\begin{adjustwidth}{-\extralength}{0cm}

\printendnotes[custom]

\reftitle{References}

\end{adjustwidth}

\begin{thebibliography}{999}

\bibitem[{Woosley} \em{et~al.}(2002){Woosley} et~al.]{Woosley:2002}
{Woosley}, S.E.;  {Heger, A.; Weaver, T.A.} 
\newblock {The evolution and explosion of massive stars}.
\newblock {\em Rev. Mod. Phys.} {\bf 2002}, {\em 74},~1015--1071.
\newblock {\url{https://doi.org/10.1103/RevModPhys.74.1015}}.

\bibitem[Janka(2016)]{Janka:2016}
Janka, H.T. Neutrino Emission from Supernovae.
\newblock In {\em Handbook of Supernovae}; Springer International Publishing:
  Cham, Switzerland, 2016; pp. 1--30.
\newblock {\url{https://doi.org/10.1007/978-3-319-20794-0_4-1}}.

\bibitem[{Mirizzi} \em{et~al.}(2016){Mirizzi} et~al.]{Mirizzi:2016}
{Mirizzi}, A.; Tamborra, I.; Janka, H.-T.; Saviano, N.; Scholberg, K.; Bollig, R.; Hüdepohl, L.; Chakraborty, S.
\newblock {Supernova neutrinos: production, oscillations and detection}.
{\em Nuovo C. Riv. Ser.} {\bf 2016}, {\em 39},~1--112.
\newblock {\url{https://doi.org/10.1393/ncr/i2016-10120-8}}.

\bibitem[Hirata \em{et~al.}(1987)Hirata et~al.]{Hirata:1987}
Hirata, K.;  Kajita, T.; Koshiba, M.; Nakahata, M.; Oyama, Y.; Sato, N.; Suzuki, A.; Takita, M.; Totsuka, Y.; Kifune, T. 
\newblock Observation of a neutrino burst from the SN1987A.
\newblock {\em Phys. Rev. Lett.} {\bf 1987}, {\em 58},~1490--1493.
\newblock {\url{https://doi.org/10.1103/PhysRevLett.58.1490}}.

\bibitem[Bionta \em{et~al.}(1987)Bionta et~al.]{Bionta:1987}
{Bionta, R.M}.;   {Blewitt, G.; Bratton, C.B.; Casper, D.; Ciocio, A.; Clauss, R.; Cortez, B.; Crouch, M.; Dye, S.T.; Errede, S.; {et~al.}} 
\newblock Observation of a neutrino burst in coincidence with supernova 1987A
  in the Large Magellanic Cloud.
\newblock {\em Phys. Rev. Lett.} {\bf 1987}, {\em 58},~1494--1496.
\newblock {\url{https://doi.org/10.1103/PhysRevLett.58.1494}}.

\bibitem[Alekseev \em{et~al.}(1987)Alekseev et~al.]{Alekseev:1987}
Alekseev, E.N.; Volchenko, V.I.
\newblock {Possible Detection of a Neutrino Signal on 23 February 1987 at the
  Baksan Underground Scintillation Telescope of the Institute of Nuclear
  Research}.
\newblock {\em JETP Lett.} {\bf 1987}, {\em 45},~589--592.

\bibitem[Aglietta \em{et~al.}(1987)Aglietta et~al.]{Aglietta:1987}
Aglietta, M.;  Badino, G.; Bologna, G.; Castagnoli, C.; Castellina, A.; Dadykin, V.L.; Fulgione, W.; Galeott, P.; Kalchukov, F.F.; Kortchaguin, B.
\newblock On the Event Observed in the Mont Blanc Underground Neutrino
  Observatory.
\newblock {\em  Europhys. Lett.} {\bf 1987}, {\em 3}, 1315.
\newblock {\url{https://doi.org/10.1209/0295-5075/3/12/011}}.

\bibitem[{H{\"u}depohl} \em{et~al.}(2010){H{\"u}depohl} et~al.]{2010:Hudepohl}
{H{\"u}depohl}, L.; Müller, B.; Janka, H.-T.; Marek, A.; Raffelt, G.G.
\newblock {Neutrino Signal of Electron-Capture Supernovae from Core Collapse to
  Cooling}.
{\em Phys. Rev. Lett.} {\bf 2010}, {\em 104},~251101.
  {\url{https://doi.org/10.1103/PhysRevLett.104.251101}}.

\bibitem[Keil \em{et~al.}(2003)Keil et~al.]{Keil:2003}
Keil, M.T.;  Raffelt, G.G.; Janka, H.-T.
\newblock {Monte Carlo study of supernova neutrino spectra formation}.
\newblock {\em Astrophys. J.} {\bf 2003}, {\em 590},~971--991,
 
\newblock {\url{https://doi.org/10.1086/375130}}.

\bibitem[O'Connor \em{et~al.}(2018)O'Connor et~al.]{OConnor:2018}
O'Connor, E.;  Bollig, R.; Burrows, A.; Couch, S.; Fischer, T.; Janka, H.-T.; Kotake, K.; Lentz, E.J.; Liebendörfer, M.; Messer, O.E.B.
\newblock Global comparison of core-collapse supernova simulations in spherical
  symmetry.
\newblock {\em J. Phys. G} {\bf 2018}, {\em 45},~104001.
\newblock {\url{https://doi.org/10.1088/1361-6471/aadeae}}.

\bibitem[{Burrows} and {Vartanyan}(2021)]{Burrows:2021}
{Burrows}, A.; {Vartanyan}, D.
\newblock {Core-collapse supernova explosion theory}.
\newblock {\em Nature} {\bf 2021}, {\em 589},~29--39.
 {\url{https://doi.org/10.1038/s41586-020-03059-w}}.

\bibitem[{von Krosigk} \em{et~al.}(2013){von Krosigk} et~al.]{vonkrosigk:2013}
{Von Krosigk}, B.;  Neumann, L.; Nolte, R.; Röttger, S.; Zuber, K.
\newblock {Measurement of the proton light response of various LAB based
  scintillators and its implication for supernova neutrino detection via
  neutrino-proton scattering}.
\newblock {\em Eur. Phys. J. C} {\bf 2013}, {\em 73},~2390.
\newblock {\url{https://doi.org/10.1140/epjc/s10052-013-2390-1}}.

\bibitem[{Li} \em{et~al.}(2019){Li} et~al.]{Li:2019}
{Li}, H.L.;  Huang, X.; Li, Yu.; Wen, Li.; Zhou, S.
\newblock {Model-independent approach to the reconstruction of multiflavor
  supernova neutrino energy spectra}. {\em Phys. Rev. D} {\bf 2019}, {\em 99},~123009.
  {\url{https://doi.org/10.1103/PhysRevD.99.123009}}.

\bibitem[{Lunardini} \em{et~al.}(2001){Lunardini} et~al.]{Lunardini:2001v2}
{Lunardini}, C.;  Smirnov, A.Y.
\newblock {Supernova neutrinos: Earth matter effects and neutrino mass
  spectrum}.
\newblock {\em Nucl. Phys.  D} {\bf 2001}, {\em 616}, 307--348.
 {\url{https://doi.org/10.1016/S0550-3213(01)00468-0}}.

\bibitem[{Lunardini} \em{et~al.}(2003){Lunardini} et~al.]{Lunardini:2003}
{Lunardini}, C.;  Smirnov, A.Y.
\newblock {Probing the neutrino mass hierarchy and the 13-mixing with
  supernovae}.
\newblock {\em J. Cosmol. Astropart. Phys.} {\bf 2003}, {\em 2003},~9.
 {\url{https://doi.org/10.1088/1475-7516/2003/06/009}}.

\bibitem[{Capozzi} \em{et~al.}(2018){Capozzi} et~al.]{Capozzi:2018}
{Capozzi}, F.; Dasgupta, B.; Mirizzi, A.
\newblock {Model-independent diagnostic of self-induced spectral equalization
  versus ordinary matter effects in supernova neutrinos}.
\newblock {\em Phys. Rev. D} {\bf 2018}, {\em 98},~063013.
  {\url{https://doi.org/10.1103/PhysRevD.98.063013}}.

\bibitem[{Tamborra} \em{et~al.}(2012){Tamborra} et~al.]{Tamborra:2012}
{Tamborra}, I.; Raffelt, G.G.; Hüdepohl, L.; Janka, Ha.
\newblock {Impact of eV-mass sterile neutrinos on neutrino-driven supernova
  outflows}.
\newblock {\em J. Cosmol. Astropart. Phys.} {\bf 2012}, {\em 2012},~13.
 {\url{https://doi.org/10.1088/1475-7516/2012/01/013}}.

\bibitem[{Giunti} and {Laveder}(2003)]{Giunti:2003}
{Giunti}, C.; {Laveder}, M.
\newblock {Neutrino Mixing}.
\newblock {\em arXiv} {\bf 2003},  {arXiv:hep-ph/0310238}. 

\bibitem[{Esteban} \em{et~al.}(2020){Esteban} et~al.]{Esteban:2020}
{Esteban}, I.;  Gonzalez-Garcia, M.C.; Maltoni, M.; Schwetz, T.; Zhou, A.
\newblock {The fate of hints: updated global analysis of three-flavor neutrino
  oscillations}.
\newblock {\em Int. J. High Energy Phys.} {\bf 2020}, {\em 2020},~178.
  {\url{https://doi.org/10.1007/JHEP09(2020)178}}.

\bibitem[Kajita(2010)]{Kajita:2010}
Kajita, T.
\newblock {Atmospheric Neutrinos and Discovery of Neutrino Oscillations}.
\newblock {\em Proc. Japan Acad. B} {\bf 2010}, {\em 86},~303--321.
\newblock {\url{https://doi.org/10.2183/pjab.86.303}}.

\bibitem[{Scholberg}(2018)]{Scholberg:2018}
{Scholberg}, K.
\newblock {Supernova signatures of neutrino mass ordering}.
\newblock {\em J. Phys. Nucl. Phys.} {\bf 2018}, {\em
  45},~014002.
   {\url{https://doi.org/10.1088/1361-6471/aa97be}}.

\bibitem[{Brdar} and {Xu}(2022)]{Brdar:2022}
{Brdar}, V.; {Xu}, X.J.
\newblock {Timing and multi-channel: novel method for determining the neutrino
  mass ordering from supernovae}.
\newblock {\em J. Cosmol. Astropart. Phys.} {\bf 2022}, {\em 2022},~67.
  {\url{https://doi.org/10.1088/1475-7516/2022/08/067}}.

\bibitem[{Jes{\'u}s-Valls}(2022)]{Valls:2022}
{Jes{\'u}s-Valls}, C.
\newblock {Uncovering the neutrino mass ordering with the next galactic
  core-collapse supernova neutrino burst}.
\newblock {\em arXiv} {\bf 2022}, arXiv:2210.11676,
 

\bibitem[{Balantekin} and {Y{\"u}ksel}(2005)]{Balantekin:2004}
{Balantekin}, A.B.; {Y{\"u}ksel}, H.
\newblock {Neutrino mixing and nucleosynthesis in core-collapse supernovae}.
\newblock {\em New J. Phys.} {\bf 2005}, {\em 7},~51.
 {\url{https://doi.org/10.1088/1367-2630/7/1/051}}.

\bibitem[Sarikas \em{et~al.}(2012)Sarikas, Raffelt, Hudepohl, and
  Janka]{Sarikas:2011}
Sarikas, S.; Raffelt, G.G.; Hudepohl, L.; Janka, H.T.
\newblock {Suppression of Self-Induced Flavor Conversion in the Supernova
  Accretion Phase}.
\newblock {\em Phys. Rev. Lett.} {\bf 2012}, {\em 108},~061101.
{\url{https://doi.org/10.1103/PhysRevLett.108.061101}}.

\bibitem[{Chakraborty} \em{et~al.}(2011){Chakraborty}, {Fischer}, {Mirizzi},
  {Saviano}, and {Tom{\`a}s}]{Chakraborty:2011}
{Chakraborty}, S.; {Fischer}, T.; {Mirizzi}, A.; {Saviano}, N.; {Tom{\`a}s}, R.
\newblock {No Collective Neutrino Flavor Conversions during the Supernova
  Accretion Phase}.
\newblock {\em Phys. Rev. Lett.} {\bf 2011}, {\em 107},~151101.
 {\url{https://doi.org/10.1103/PhysRevLett.107.151101}}.

\bibitem[Dighe and Smirnov(2000)]{Dighe:2000}
Dighe, A.S.; Smirnov, A.Y.
\newblock {Identifying the neutrino mass spectrum from the neutrino burst from
  a supernova}.
\newblock {\em Phys. Rev. D} {\bf 2000}, {\em 62},~033007.
 {\url{https://doi.org/10.1103/PhysRevD.62.033007}}.

\bibitem[Cottingham \em{et~al.}(2001)Cottingham, Greenwood, and
  Greenwood]{cottingham:2001}
Cottingham, W.; Greenwood, D.; Greenwood, D.
\newblock {\em An Introduction to Nuclear Physics}; Cambridge University Press: Cambridge, UK, 2001.

\bibitem[Zyla \em{et~al.}(2020)Zyla et~al.]{Zyla:2020}
Zyla, A. et al. [Particle Data Group]
\newblock {Review of Particle Physics}.
\newblock {\em Prog. Theor. Exp. Phys.} {\bf 2020}, {\em 2020},~083C01.
{\url{https://doi.org/10.1093/ptep/ptaa104}}.

\bibitem[Dasgupta and Beacom(2011)]{Dasgupta:2011}
Dasgupta, B.; Beacom, J.F.
\newblock {Reconstruction of supernova $\nu_\mu$, $\nu_\tau$, anti-$\nu_\mu$,
  and anti-$\nu_\tau$ neutrino spectra at scintillator detectors}.
\newblock {\em Phys. Rev. D} {\bf 2011}, {\em 83},~113006.
  {\url{https://doi.org/10.1103/PhysRevD.83.113006}}.

\bibitem[Sibley(2015)]{SNOplus:2015}
Sibley, L.
\newblock {SNO+: Physics program and status update}.
\newblock {\em AIP Conf. Proc.} {\bf 2015}, {\em 1604},~449--455.
\newblock {\url{https://doi.org/10.1063/1.4883464}}.

\bibitem[{V{\"a}{\"a}n{\"a}nen} and {Volpe}(2011)]{Volpe:2011}
{V{\"a}{\"a}n{\"a}nen}, D.; {Volpe}, C.
\newblock {The neutrino signal at HALO: learning about the primary supernova
  neutrino fluxes and neutrino properties}.
\newblock {\em J. Cosmol. Astropart. Phys.} {\bf 2011}, {\em 2011},~019.
 {\url{https://doi.org/10.1088/1475-7516/2011/10/019}}.

\bibitem[Engel \em{et~al.}(2003)Engel, McLaughlin, and Volpe]{Engel:2003}
E {ngel, J.; McLaughlin, G.C.; Volpe, C.} 
\newblock What can be learned with a lead-based supernova-neutrino detector?
\newblock {\em Phys. Rev. D} {\bf 2003}, {\em 67},~013005.
\newblock {\url{https://doi.org/10.1103/PhysRevD.67.013005}}.

\bibitem[{Scholberg} \em{et~al.}(2021){Scholberg}, {Albert}, and
  {Vasel}]{Scholberg:2021}
{Scholberg}, K.; {Albert}, J.B.; {Vasel}, J.
\newblock {SNOwGLoBES: SuperNova Observatories with GLoBES}.
\newblock  {Astrophysics Source Code Library, record ascl:2109.019   
},  2021. \url{https://ascl.net/2109.019}.

\bibitem[{Abi} \em{et~al.}(2021){Abi} et~al.]{Abi:2021}
{Abi}, B.;  Acciarri, R.; Acero, M.A.; Adamov, G.; Adams, D.; Adinolfi, M.; Ahmad, Z.; Ahmed, J.; Alion, T.; Monsalve, S.A.
\newblock {Supernova neutrino burst detection with the deep underground
  neutrino experiment}.
\newblock {\em Eur. Phys. J. C} {\bf 2021}, {\em 81},~423.
{\url{https://doi.org/10.1140/epjc/s10052-021-09166-w}}.

\bibitem[{Abi} \em{et~al.}(2020){Abi} et~al.]{Abi:2020-TDR}
{Abi}, B.; Acciarri, R.; Acero, M.A.; Adamov, G.; Adams, D.; Adinolfi, M.; Ahmad, Z.; Ahmed, J.; Alion, T.; Monsalve, S.A.
\newblock {Volume I. Introduction to DUNE}.
\newblock {\em J. Instrum.} {\bf 2020}, {\em 15},~T08008.
{\url{https://doi.org/10.1088/1748-0221/15/08/T08008}}.

\bibitem[{DUNE Collaboration} \em{et~al.}(2020){DUNE Collaboration}, {Abi},
  et~al.]{Abi:2020}
{Abi}, B.  et~al. [DUNE Collaboration]
\newblock {Prospects for Beyond the Standard Model Physics Searches at the Deep
  Underground Neutrino Experiment}.
\newblock {\em arXiv} {\bf 2020}, arXiv:2008.12769.

\bibitem[Gardiner(2021)]{Gardiner:2021}
Gardiner, S.
\newblock {Simulating low-energy neutrino interactions with MARLEY}.
\newblock {\em Comput. Phys. Commun.} {\bf 2021}, {\em 269},~108123.
{\url{https://doi.org/10.1016/j.cpc.2021.108123}}.

\bibitem[Barger \em{et~al.}(2002)Barger, Marfatia, and Wood]{Barger:2001}
Barger, V.; Marfatia, D.; Wood, B.P.
\newblock {Inverting a supernova: Neutrino mixing, temperatures and binding
  energy}.
\newblock {\em Phys. Lett. B} {\bf 2002}, {\em 547},~37--42.
  {\url{https://doi.org/10.1016/S0370-2693(02)02741-7}}.

\bibitem[Minakata \em{et~al.}(2002)Minakata, Nunokawa, Tomas, and
  Valle]{Minakata:2001}
Minakata, H.; Nunokawa, H.; Tomas, R.; Valle, J.W.F.
\newblock {Probing supernova physics with neutrino oscillations}.
\newblock {\em Phys. Lett. B} {\bf 2002}, {\em 542},~239--244.
 {\url{https://doi.org/10.1016/S0370-2693(02)02376-6}}.

\bibitem[Gallo~Rosso \em{et~al.}(2018)Gallo~Rosso, Vissani, and
  Volpe]{GalloRosso:2017}
Gallo~Rosso, A.; Vissani, F.; Volpe, M.C.
\newblock {What can we learn on supernova neutrino spectra with water Cherenkov
  detectors?}
\newblock {\em J. Cosmol. Astropart. Phys.} {\bf 2018}, {\em 04},~040.
  {\url{https://doi.org/10.1088/1475-7516/2018/04/040}}.

\bibitem[Engel \em{et~al.}(2003)Engel, McLaughlin, and Volpe]{Engel:2002}
E {ngel, J.; McLaughlin, G.C.; Volpe, C.}
\newblock {What can be learned with a lead based supernova neutrino detector?}
\newblock {\em Phys. Rev. D} {\bf 2003}, {\em 67},~013005.
 {\url{https://doi.org/10.1103/PhysRevD.67.013005}}.

\bibitem[Stringer(2019)]{Stringer:2019ztj}
Stringer, M.
\newblock {Sensitivity of SNO+ to Supernova Neutrinos}.
\newblock Ph.D. Thesis, Sussex University:  {Sussex, UK,}  2019.

\bibitem[Dembinski and et~al.(2020)]{iminuit}
Dembinski, H.; Ongmongkolkul; Deil, P.; Schreiner, C.; Feickert, H.; Burr, M.; Watson, C.; Rost, J.; Pearce, F.; Geiger, A.; et al.
\newblock  {scikit-hep/iminuit, v2.24.0, Zenodo} 
 {\bf 2020}.
\newblock {\url{https://doi.org/10.5281/zenodo.8249703}}.

\bibitem[James and Roos(1975)]{James:1975dr}
James, F.; Roos, M.
\newblock {Minuit: A System for Function Minimization and Analysis of the
  Parameter Errors and Correlations}.
\newblock {\em Comput. Phys. Commun.} {\bf 1975}, {\em 10},~343--367.
\newblock {\url{https://doi.org/10.1016/0010-4655(75)90039-9}}.

\bibitem[Wolfenstein(1978)]{Wolfenstein:1977}
 {Wolfenstein, L.} 
\newblock {Neutrino Oscillations in Matter}.
\newblock {\em Phys. Rev.} {\bf 1978}, {\em D17},~2369--2374.
{\url{https://doi.org/10.1103/PhysRevD.17.2369}}.


\bibitem[{Strumia} and {Vissani}(2003)]{Strumia:2003}
{Strumia}, A.; {Vissani}, F.
\newblock {Precise quasielastic neutrino/nucleon cross-section}.
\newblock {\em Phys. Lett. B} {\bf 2003}, {\em 564},~42--54,
{\url{https://doi.org/10.1016/S0370-2693(03)00616-6}}.

\bibitem[Weinberg(1972)]{Weinberg:1972}
Weinberg, S.
\newblock Effects of a Neutral Intermediate Boson in Semileptonic Processes.
\newblock {\em Phys. Rev. D} {\bf 1972}, {\em 5},~1412--1417.
\newblock {\url{https://doi.org/10.1103/PhysRevD.5.1412}}.

\bibitem[Ahrens \em{et~al.}(1987)Ahrens et~al.]{Ahrens:1987}
 {Ahrens, L.A}.; Aronson, S. H.; Connolly, P. L.; Gibbard, B. G.; Murtagh, M. J.; Murtagh, S. J.; Terada, S.; White, D. H.; Callas, J. L.; Cutts, D.;  et~al. 
\newblock Measurement of neutrino-proton and antineutrino-proton elastic
  scattering.
\newblock {\em Phys. Rev. D} {\bf 1987}, {\em 35},~785--809.
\newblock {\url{https://doi.org/10.1103/PhysRevD.35.785}}.

\bibitem[Beringer \em{et~al.}(2012)Beringer et~al.]{Beringer:2012}
 {Beringer, J.;  Arguin J. F.; Barnett, R. M.; Copic, K.; Dahl, O.; Groom, D. E.; Lin, C. J.; Lys, J.; Murayama, H.; Wohl, C. G.;  et~al.} 
\newblock Review of Particle Physics.
\newblock {\em Phys. Rev. D} {\bf 2012}, {\em 86},~010001.
\newblock {\url{https://doi.org/10.1103/PhysRevD.86.010001}}.

\bibitem[Armbruster \em{et~al.}(1998)Armbruster et~al.]{Armbruster:1998}
Armbruster, B.  et~al. [KARMEN Collaboration].
\newblock Measurement of the weak neutral current excitation
  12C($\nu\mu\nu\mu'$)12C$^{*}(1+,1;15.1MeV)$ at $E_{\nu\mu}=$29.8 MeV.
\newblock {\em Physics Letters B} {\bf 1998}, {\em 423},~15--20.
{\url{https://doi.org/10.1016/S0370-2693(98)00087-2}}.

\bibitem[Donnelly and Peccei(1979)]{Donnelly:1979}
Donnelly, T.; Peccei, R.
\newblock Neutral current effects in nuclei.
\newblock {\em Phys. Rep.} {\bf 1979}, {\em 50},~1--85.
{\url{https://doi.org/https://doi.org/10.1016/0370-1573(79)90010-3}}.

\bibitem[{Bhattacharyya} and {Dasgupta}(2022)]{Bhattacharyya:2022}
{Bhattacharyya}, S.; {Dasgupta}, B.
\newblock {Elaborating the ultimate fate of fast collective neutrino flavor
  oscillations}.
\newblock {\em \prd} {\bf 2022}, {\em 106},~103039.
{\url{https://doi.org/10.1103/PhysRevD.106.103039}}.

\bibitem[{Richers} \em{et~al.}(2021){Richers}, {Willcox}, {Ford}, and
  {Myers}]{Richers:2021}
{Richers}, S.; {Willcox}, D.E.; {Ford}, N.M.; {Myers}, A.
\newblock {Particle-in-cell simulation of the neutrino fast flavor
  instability}.
\newblock {\em \prd} {\bf 2021}, {\em 103},~083013.
 {\url{https://doi.org/10.1103/PhysRevD.103.083013}}.

\bibitem[{Wu} \em{et~al.}(2021){Wu}, {George}, {Lin}, and {Xiong}]{Wu:2021}
{Wu}, M.R.; {George}, M.; {Lin}, C.Y.; {Xiong}, Z.
\newblock {Collective fast neutrino flavor conversions in a 1D box: Initial
  conditions and long-term evolution}.
\newblock {\em \prd} {\bf 2021}, {\em 104},~103003.
 {\url{https://doi.org/10.1103/PhysRevD.104.103003}}.

\bibitem[{Dasgupta} and {Mirizzi}(2015)]{Dasgupta:2015}
{Dasgupta}, B.; {Mirizzi}, A.
\newblock {Temporal instability enables neutrino flavor conversions deep inside
  supernovae}.
\newblock {\em \prd} {\bf 2015}, {\em 92},~125030.
  {\url{https://doi.org/10.1103/PhysRevD.92.125030}}.

\bibitem[{Sawyer}(2016)]{Sawyer:2016}
{Sawyer}, R.F.
\newblock {Neutrino Cloud Instabilities Just above the Neutrino Sphere of a
  Supernova}.
\newblock {\em \prl} {\bf 2016}, {\em 116},~081101.
 {\url{https://doi.org/10.1103/PhysRevLett.116.081101}}.

\bibitem[{Dasgupta} \em{et~al.}(2017){Dasgupta}, {Mirizzi}, and
  {Sen}]{Dasgupta:2017}
{Dasgupta}, B.; {Mirizzi}, A.; {Sen}, M.
\newblock {Fast neutrino flavor conversions near the supernova core with
  realistic flavor-dependent angular distributions}.
\newblock {\em \jcap} {\bf 2017}, {\em 2017},~019.
{\url{https://doi.org/10.1088/1475-7516/2017/02/019}}.

\end{thebibliography}
\end{document}